\documentclass[10pt]{article}
\usepackage[margin=1.0in]{geometry}
\usepackage[utf8]{inputenc}
\usepackage[english]{babel}
\usepackage[numbers]{natbib}
\usepackage{graphicx}
\usepackage{framed}
\usepackage{xurl}
\usepackage{hyperref}
\usepackage{amsmath}
\usepackage{amsfonts}
\usepackage{enumitem}
\usepackage{xcolor}
\usepackage{colortbl}
\usepackage{booktabs}
\usepackage{comment}
\usepackage{titlesec}
\usepackage{multicol}
\definecolor{airforceblue}{rgb}{0.36, 0.54, 0.66}
\definecolor{bostonuniversityred}{rgb}{0.8, 0.0, 0.0}
\definecolor{cadmiumgreen}{rgb}{0.0, 0.42, 0.24}
\usepackage{wrapfig}
\usepackage[justification=centering, font=footnotesize]{caption}

\newcommand{\R}{\mathbb{R}}

\title{Operational and Economy-Wide Impacts of Compound Cyberattacks and Extreme Weather Events on Electric Power Networks}
\author{Charalampos Avraam, Luis Ceferino, Yury Dvorkin}
\date{}

\begin{document}

\maketitle

\section*{Abstract}

The growing frequencies of extreme weather events and cyberattacks give rise to a novel threat where a malicious cyber actor aims to disrupt stressed components of critical infrastructure systems immediately before, during, or shortly after an extreme weather event. In this paper, we initiate the study of Compound Cyber-Physical Threats and develop a two-stage framework for the analysis of operational disruptions in electric power networks and economy-wide impacts under three scenarios: a Heatwave, a Cyberattack, and a Compound scenario when the Cyberattack is timed with the Heatwave. In the first stage, we use a bilevel optimization problem to represent the adversarial rationale of a cyberattacker in the upper level. In the lower level, we model disruptions in the electric power network using an optimal power flow model. In the second stage, we couple the disruption of electricity supply with a Computable General Equilibrium model to elucidate the impacts on all economic sectors. For the New York Independent System Operator, we find that a 9\% demand increase in a Heatwave may not lead to unserved load. The Cyberattack can lead to 4\% of unserved electric load in Long Island, while the Compound scenario can increase unserved electric load in Long Island to 13\% and affect almost 600,000 customers. Our results show that the activity of state and local government enterprises can decrease by 30\% in the Compound scenario. We conclude that the vulnerability of federal, state, and local government enterprises to electricity disruptions can affect a broad range of populations.

\paragraph{Keywords} heatwave, cyberattack, compound risks, CGE, bilevel optimization, cyber-physical systems

\section{Introduction} 

Information communication and control (ICC) technologies enhance the efficiency and flexibility of infrastructure operations \citep{colombo2017}, but expose critical infrastructure components to cyberattacks. In infrastructure systems with strong coupling of ICC technologies and physical operations, malware infiltration can result in large disruptions 
\citep{mclaughlin2016}. According to the Industrial Control Systems Cyber Emergency Response Team, the annual number of data breaches and exposed records in the U.S. grew sixfold and twofold,  respectively, between 2005 and 2020.
\citep{ics-cert2015}. 
Across all economic sectors, cyberattacks increase in frequency and intensity \citep{Granato2019}, but energy infrastructure is at particular high risk. Energy was the second most targeted economic sector in 2015 \citep{ics-cert2015}. 

Cyberattacks can be more likely and damaging when infrastructures are facing stresses than in normal operating conditions. For example, the U.S. Federal Bureau of Investigation noted in 2022 that cyberattackers may target U.S. food infrastructure more aggressively during the critical planting and harvest seasons, \textit{e.g.,} February-March and September-October for grains \cite{fbi2022}. Since the surge of COVID-19, data breaches against healthcare organizations have been persistently increasing, compared to 2018, and affected a record-high of 45 million individuals in 2021 \cite{ci2022}. The more recent attack against oil terminals in Germany and in the Amsterdam-Rotterdam-Antwerp area in late January and early February 2022 \cite{wef2022} happened during Europe's peak gas demand season \citep{oies2020}. 

Extreme weather events also damage critical infrastructures. For example, they damage electric power substations, utility poles, and distribution lines \cite{CeferinoBayesianPanel2021,CeferinoDERWCEE2020,patel2021}. Damaged electric power components can disrupt drinking water pumps, sewage pump stations, telecommunications cellular tower sites,
causing substantial economy-wide financial losses. \textit{Hurricane Isaias} left almost 800,000 utility customers without power in New York in 2020 \citep{latto2021} and \textit{Hurricane Ida} more than 1.2 million customers in Louisiana in 2021 \citep{beven2022}, largely due to damaged utility poles. A substation explosion caused by \textit{Hurricane Sandy} left most of Lower Manhattan without power for more than four consecutive days in 2012 \citep{nyc2013b}, affecting also the transportation and telecommunications infrastructure systems \citep{usdc2013}.

Disruptions in the electric power sector can also propagate to  food infrastructure systems, \textit{e.g.,} in the case of \textit{Winter Storm Jonas} in 2016 \cite{biehl2017}. In addition, disruptions in the electric power sector propagate across sectors and cause economy-wide losses. Estimates of the economy-wide losses of the 2003 Northeastern Blackout range between \$4-10 billion \citep{usdoe2004}.
Nationally, the 10-year economic losses in the US due to extreme weather events have quadrupled between 1980-2020, averaging \$120 billion per year in 2016-2020  \citep{smith2021}.

The growing frequency of extreme weather events and cyberattacks manifest a novel threat where a malicious cyber actor disrupts stressed components of critical infrastructure immediately before, during, or shortly after an extreme weather event. Therefore, this paper aims to initiate the investigation of compound cyberattacks and extreme weather events, hereinafter referred to as \textit{Compound Cyber-Physical Threats}, in critical infrastructure systems. We focus on the electric power sector due to its importance for the operation of other critical infrastructure systems and its vulnerability to cyberattacks and extreme weather events.  Recognizing the potential of \textit{Compound Cyber-Physical Threats}, the U.S. National Guard in coordination with the City of Houston in 2018 and the State of Indiana in 2021 designed and conducted three-day drills of a cyberattack striking during an extreme weather event \citep{houston-mayor2018,iecc2021}. However, there does not exist a framework to (a) understand the compounding effect of a cyberattack timed with an extreme weather event compared to a cyberattack under normal operating conditions, (b) identify the spatial distribution of critically-vulnerable infrastructure components within an electric power network, and (c) generate realistic, synthetic scenarios of \textit{Compound Cyber-Physical Threats} for service disruptions and their impacts on the economy. Thus, this paper investigates the following research questions:
\begin{itemize}

    \item What is the disruption potential of \textit{Compound Cyber-Physical Threats} for electric power networks?
    
    \item Which regional electric power  network components are more vulnerable to a cyberattack timed with an operational disruption caused by an extreme weather event?   
    
    \item How do operational disruptions caused by a \textit{Compound Cyber-Physical Threat} against electric power network components affect the activity of other sectors in the economy?

\end{itemize}

In this paper, we identify vulnerable  electric power  network components across New York State and within New York City, and assess economy-wide 
operational and price disruptions due to \textit{Compound Cyber-Physical Threats} across sectors. We focus on New York City because it is the most densely populated city in the U.S. \citep{uscb2021}, hence a regional disruption of electricity supply can impact a larger population and cause more severe cascading failures to interdependent infrastructure systems compared to other cities. Moreover, the New York State is the most  exposed U.S. state to physical and cyber risks according to Lloyd's and the Cambridge Centre for Risk Studies \citep{cjbs2015} that measure the impact of all risks on the Gross Domestic Product (GDP@Risk in Billion US\$).

The assessment of the operational disruption of a \textit{Compound Cyber-Physical Threat} requires integrating
an electric power network model capturing the physical and economic principles of the electric power network
operations in New York State, with a cyberattacker model, capturing the adversarial rationale of the
cyberattacker. We use an Optimal Power Flow model (OPF) model, which is representative of operational
practices in the New York Independent System Operator (NYISO) electric power network, to capture the response of the system operator
\citep{eldridge2017} to a heatwave. 
We focus on heatwaves because they stress the grid by increasing electricity demand. 
Heatwaves will be exacerbated by global warming \citep{romitti2022}, rendering them more costly \citep{cohen2018} and dangerous for vulnerable populations \citep{klinenberg1999, cong2020} in future climates. A bilevel optimization problem models the adversarial rationale of a cyberattacker anticipating the system
operator’s response \citep{castillo2019, karangelos2022}. 

To further assess the economy-wide
impacts, we couple the cyberattacker bilevel optimization problem with a Computable General Equilibrium
model (CGE), which can capture flow of money and goods across all sectors in an economy \citep{rutherford2019}. The assessment of sectoral disruptions caused by natural hazards \citep{rose2004}, terrorist attacks \citep{rose2007} and cyberattacks \citep{dreyer2018, eling2022} in existing literature is exogenous and is based on a series of assumptions. For example, Rose \textit{et al.,} \citep{rose2007}, assume that completely disconnecting the transmission lines between Los Angeles County and the rest of the electric power network results in 100\% unserved load for two weeks. The authors then constrain electricity supply in a CGE to simulate the response of all other sectors. Instead, in our framework, unserved load is the result of the bilevel optimization problem. Following a failure, our framework considers the physical response of the electric power network and the decisions of the system operator to derive unserved load. For example, in an attack against transmission lines in our framework, the system operator may choose to operate the electric power network in islanded mode to support parts of the network using local generation and avoid a complete blackout. 
Then, we inform the constraints on regional electricity supply in a CGE in Rose \textit{et al.,} \citep{rose2007} based on the results of the bilevel optimization problem. Hence, the coupling extends the existing literature by relaxing the assumptions on the operational disruptions of natural hazards and cyberattacks. 
We do not aim to quantify disaster impacts in the economy, as in Markhvida \textit{et al.,} \citep{markhvida2020}, but understand the response of all economic sectors to operational disruptions in the electric power network.

The rest of the paper is organized as follows. Section \ref{sec:scenario-design} describes the assumptions behind the \textit{Compound Cyber-Physical Threat} scenarios and Section \ref{sec:methodology} introduces the proposed modeling framework and methodology. We identify electric power network and inter-sectoral outcomes under combinations of extreme weather events and cyberattacks in Section \ref{sec:results} and discuss resilience implications and limitations in Section \ref{sec:discussion}. We conclude in Section \ref{sec:conclusions} with the summary of the main findings and future research.

\section{Scenario Design}\label{sec:scenario-design}

We first describe the \textit{Baseline} scenario, which will serve as a reference for the \textit{Cyberattack}, \textit{Heatwave}, and \textit{Compound Cyber-Physical Threat} scenarios. Among different extreme weather events, New York City is particularly vulnerable to \textit{Heatwaves} \citep{ramamurthy2017}. First, although winter outages result in greater welfare losses in the current climate and electrification levels, summer outages due to \textit{Heatwaves} can be more costly under warmer future climate \citep{cohen2018}. Second, higher temperatures increase power demand directly due to air conditioners \citep{bessec2008}, which in turn strains the electric power network. The heat-driven strain is particularly critical in power network which have their annual peak consumption in summer, \textit{e.g.,} the NYISO network. Third, \textit{Heatwaves}, or extreme heat events, disproportionately affect the elderly \citep{klinenberg1999}, low-income \citep{cong2020} and other vulnerable populations \citep{eckstrom2022}, Finally, densely populated cities, including New York City, can become urban heat islands during summer, \textit{i.e.,} the cities are subject to increased temperatures compared to neighboring rural areas \citep{scott2018}. In what follows, we describe the assumptions behind each scenario.

\begin{figure}[h!]
	    \centering	\includegraphics[width=1.0\textwidth]{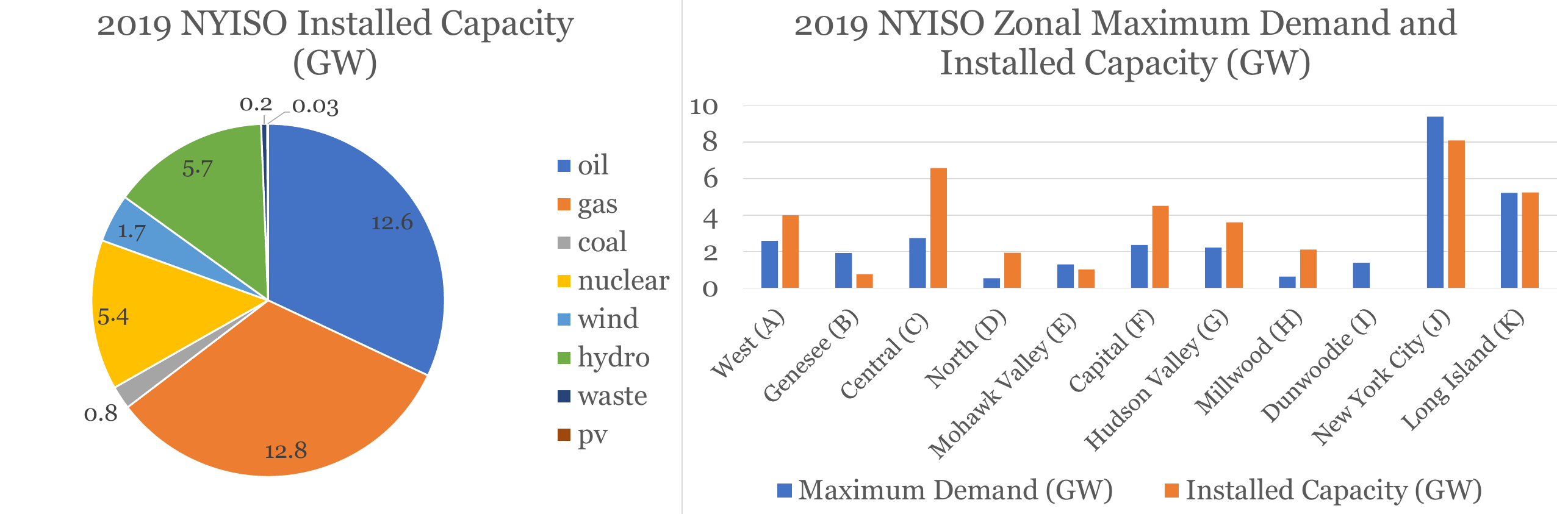}
		\caption{\label{fig:baseline-nyiso}\footnotesize \textbf{Left:} NYISO installed capacity in 2019 by generation technology. \textbf{Right:} Maximum Demand and installed power generation capacity by NYISO Zone.}
	\end{figure}

\paragraph{Baseline.} We obtain the 2019 installed capacity and electricity demand for each NYISO zone (Figure \ref{fig:baseline-nyiso}) and an hourly summer load-duration curve for NYISO zones from \citep{khan2022-risk, nyiso-goldbook2019}. 
We retrieve the capacity and impedances of transmission lines of the  NYISO electric power network from \citep{khan2022-risk}. 
To the best of our knowledge, a 2019 estimate of the number of NYISO customers is not available. Therefore, in what follows, we use a total number of customers equal to 19.8 million, reported by the Federal Energy Regulatory Commission (FERC) on July 2020 \citep{ferc2020}.

\paragraph{Heatwave.} We base the formulation of the \textit{Heatwave} scenario on the 2022 analysis by Rommitti and Wing \citep{romitti2022}. The authors study the impact of increasing temperatures on annual electricity demand by 2050 for 36 cities, including New York City, using 
projections of future temperature from 21 global climate models 
\citep{jones2016}. For New York City, they find that yearly mean peak electricity demand can increase approximately by 5.6\%, and the maximum increase can be 7.5\%. 
However, actual and predicted peak electricity demand can differ. For example,  the short-term prediction of peak electricity demand in each node of the NYISO deviated by 16\% from the actual demand in 2020 \citep{nyiso-goldbook2021}. 
Therefore, in our \textit{Heatwave} scenario, we assume that the increase in demand is 20\% higher than the maximum calculated increase in \citep{romitti2022}, \textit{i.e.,} a $7.5\cdot 1.2 = 9\%$ increase in hourly electricity demand for all NYISO zones.

\paragraph{Cyberattack.} The cyberattacker aims to maximize the operational disruption to the electric power network by compromising power generation capacity, transmission capacity, and limits on voltage angle differences. Since generation, transmission, and end-use components use Supervisory Control And Data Acquisition (SCADA)  devices and Programmable Logic Controller (PLC) interfaces \citep{acharya2020}, the cyberattacker can modify SCADA and PLC firmware \citep{p40,xwang2016tcad} and hardware \citep{confirm,Wang2016b} to execute unauthorized code. In our framework, the cyberattacker can inject spoof measurements and  modify control algorithms and messages exhanged SCADA-SCADA, SCADA-PLC, PLC-PLC communications to constrain and manipulate regional generation, flow and angle difference values and limits. The NYISO oversees transmission components and participates in the cybersecurity initiatives of the North American Electric Corporation \citep{epri2015}. Individual generators do not necessarily participate in the North American Electric Power Corporation. For that, we assume that generators are more vulnerable to cyberattacks in the default \textit{Cyberattack} scenario. Equivalently, it is significantly more difficult, \textit{i.e.,} costlier, for a cyberattacker to compromise a transmission component compared to generation capacity in the default \textit{Cyberattack} scenario where $\tilde{c}_f=5\cdot\tilde{c}^g$.  Because our knowledge is limited regarding the cyberattacker capabilities, we conduct sensitivity analysis on the available cyberattacker resources $(\tilde{b})$ and the relative difficulty of compromising transmission components versus power generation capacity $(\frac{\tilde{c}^g}{\tilde{c}^f})$.    

\paragraph{Compound Cyber-Physical Threat.} In the \textit{Compound Cyber-Physical Threat} scenario, the \textit{Cyberattack} is timed with the \textit{Heatwave}. We assume that the cyberattacker has access to heatwave advisories or predictions, which are possible up to 72 hours ahead of the event \citep{puri2013} to prepare and launch the cyberattack. Moreover, we assume that the cyberattack is perfectly timed with the heatwave to maximize unserved electricity. In this scenario, electricity demand increases by 9\% for all hours and NYISO zones, similarly to the \textit{Heatwave} scenario, and the cyberattacker compromises transmission and generation components simultaneously. Given the uncertainty regarding the cyberattacker capabilities, we conduct sensitivity analysis on the available cyberattacker resources and the relative difficulty of compromising transmission components in this scenario as well.

\section{Methods}\label{sec:methodology}
\begin{figure}[h!]
	    \centering	\includegraphics[width=0.6\textwidth]{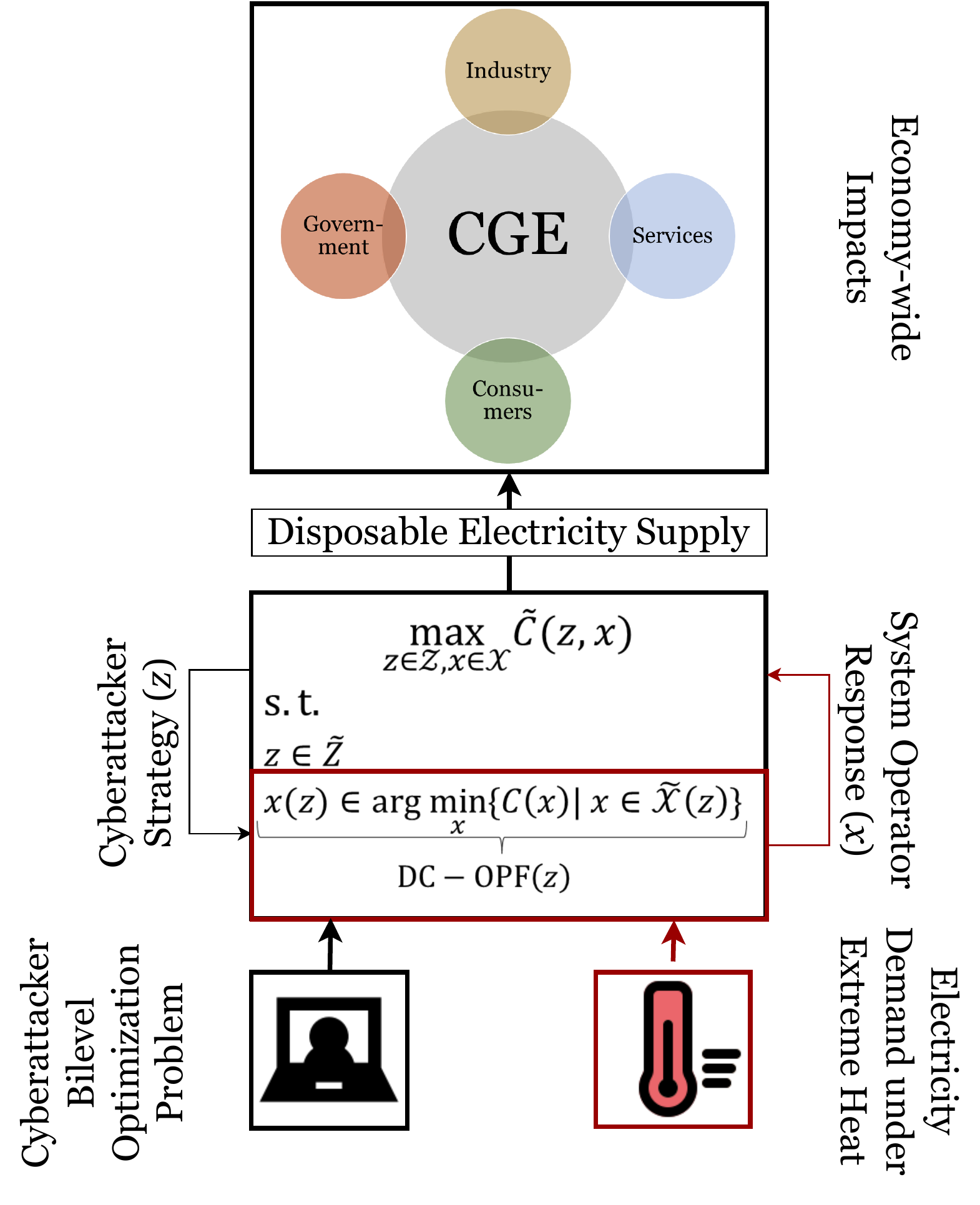}
		\caption{\label{fig:methodology}\footnotesize Proposed Framework. A \textit{Heatwave} stresses the NYISO electric power network, modeled as a DC-OPF. A cyberattacker chooses strategy $z$ which maximizes unserved load $\tilde{C}(\cdot)$ while anticipating the response of the system operator $x$. We input the percent change in regional electricity supply into a Computable General Equilibrium Model (CGE) to find activity adjustments of all other economic sectors under a disruption.}
	\end{figure}

In our framework, the cyberattacker attempts to maximize the disruption to the electric power network by compromising generation capacity, power flow capacity, or the limit on nodal voltage phase angle differences. Acharya \textit{et al.} \citep{acharya2020} show how a cyberattacker can infer features of  electricity transmission and distribution systems using publicly available data. Thus, we  assume that the cyberattacker has complete knowledge of the electric power network, including the location, capacity, moment of inertia, and type of controller of generators; the impedance of transformers and transmission lines; and the load damping constant for New York City. Under complete knowledge, the cyberattacker can design their strategy by fully anticipating the response of the NYISO and exacerbate the impact of the cyberattack. Hence, the complete knowledge assumption is true also under a worst-case scenario and  typical in cyber vulnerability assessments \cite{yanling2011,jingwen2015,castillo2019}. We represent the operation of the power system using a Direct Current Optimal Power Flow (DC-OPF) model, commonly used in power network operations, \textit{e.g.,} for electricity market dispatch \citep{eldridge2017}, contingency analysis \citep{capitanescu2008}, storage siting \citep{wogrin2014}, and transmission planning \citep{qingyu2019}. We pose the DC-OPF as an optimization problem, which regulates generation and power flow decisions within the power system to minimize the system-wide operating cost. Since the cyberattacker aims to maximize the electric power network disruption, the DC-OPF model is a helpful modeling tool for the attacker to determine and analyze its actions. Therefore, we formulate the cyberattacker problem as a \textit{bilevel optimization problem} where the upper level accounts for the  objective and constraints of the cyberattacker and the lower level includes the DC-OPF model (Figure \ref{fig:methodology}). The bilevel optimization framework allows us to capture the cyberattacker rationale and the anticipated  response of the system operator under the cyberattack and increased electricity demand due to a \textit{Heatwave}, which in turns informs the design of the attack strategy against electric power network components. 

Disruptions in the electric power network decrease electricity supply to all other sectors. We compute economy-wide losses for all sectors by coupling the bilevel cyberattacker model with WiNDC, a CGE model \citep{rutherford2019}. Using WiNDC, we find the response of all economic sectors to limited electricity supply, caused by a \textit{Heatwave}, \textit{Cyberattack}, and \textit{Compound Cyber-Physical Threat}. Coupling the cyberattacker bilevel optimization problem with the CGE allows us to estimate economy-wide losses, through changes in the Gross Domestic Product, and identify the top-five impacted sectors in each scenario. In what follows we detail the formulation of the bilevel problem, the solution algorithm, and the economy-wide model. Figure \ref{fig:methodology} details the proposed methodology.

\subsection{Lower-Level Decisions: DC-OPF Model}\label{subsec:methodology-ll-dcopf}

\paragraph{DC-OPF  as a Linear Program.}

We consider an electric power network with $G$ power generation technologies, $N$ nodes and $E$ edges/transmission lines. The DC-OPF model computes power generation $\left(g_{sh}\in\R^G\right)$, power flows $\left(f_{sh}\in\R^E\right)$, and voltage angles $\left(\theta_{sh}\in\R^N\right)$ within an electric power network for all four seasons $s\in\mathcal{S}$ and hours $h\in\mathcal{H}$ of representative days. Through the DC-OPF, we can also compute wholesale electricity prices $\left(\pi^d_{sh}\in\R^N\right)$ and other payments to power generators, \textit{e.g.,} rent $\left(\bar{\rho}^g_{sh}\in\R^G\right)$ when a generator is at capacity. Therefore, emergency response decisions within a DC-OPF model under a disruption include change in power generation, change in power flows, change in voltage angles $\left(\theta_{sh}\right)$, change in electricity prices (\$/kWh), and unserved electricity demand $\left(u_{sh}\in\R^N\right)$ when generation and transmission capabilities can not cover electricity demand at a node in season $s$ and hour $h$. Equation \eqref{eq:opf-lp} describes the DC-OPF model, where variables $\pi^f,\underline{\rho}^g,\bar{\rho}^g,\underline{\rho}^f,\bar{\rho}^f,\underline{\rho}^\theta,\bar{\rho}^\theta,\underline{\rho}^u_{sh},\bar{\rho}^u_{sh},\delta$ are the dual variables of the respective constraints and are given in parentheses. Table \ref{table:nomenclature} in the Supplementary Material 
provides a detailed description of all lower-level parameters and variables.

\begin{subequations}\label{eq:opf-lp}
    \begin{alignat}{5}
        &\multispan3{%
        $\begin{array}{ll}
             \min\limits_{\begin{array}{l}
         g_{sh}\in\R^G,f_{sh}\in\R^E, \\
         u_{sh},\theta_{sh}\in\R^N
         \end{array}} & C\left(g,u\right) 
        \end{array}$}\nonumber   \\
        \label{eq:opf-lp-obj}\text{s.t.} & && \\
        \label{eq:opf-lp-mkt-clrng-cnstr}& \sum_{t\in\mathcal{T}}g_{sh} + u_{sh} - d_{sh} + A^T f_{sh} = 0, & \quad\forall s\in\mathcal{S},&\quad\forall h\in\mathcal{H} & \qquad\left(\pi^d_{sh}\right)\\ 
        \label{eq:opf-lp-pow-flow-cnstr}& f_{sh} + B A \theta_{sh} = 0, & \quad\forall s\in\mathcal{S},&\quad\forall h\in\mathcal{H} & \qquad\left(\pi^f_{sh}\right) \\
        \label{eq:opf-lp-gen-lim-cnstrs}& \underline{g}_{sh} \leq g_{sh} \,\,\,\leq \bar{g}_{sh}, & \quad\forall s\in\mathcal{S},&\quad\forall h\in\mathcal{H} & \qquad\left(\underline{\rho}^g_{sh},\bar{\rho}^g_{sh}\right) \\
        \label{eq:opf-lp-flow-lim-cnstrs}& \underline{f}_{sh} \leq f_{sh} \,\,\,\leq \bar{f}_{sh}, & \quad\forall s\in\mathcal{S},&\quad\forall h\in\mathcal{H} & \qquad\left(\underline{\rho}^f_{sh},\bar{\rho}^f_{sh}\right) \\
        \label{eq:opf-lp-angle-lim-cnstrs}& \underline{\theta}_{sh} \leq A\theta_{sh} \leq \bar{\theta}_{sh}, & \quad\forall s\in\mathcal{S},&\quad\forall h\in\mathcal{H} & \qquad\left(\underline{\rho}^\theta_{sh},\bar{\rho}^\theta_{sh}\right) \\
        \label{eq:opf-lp-unsd-lim-cnstrs}& 0\,\,\, \leq u_{sh} \,\,\,\,\leq d_{sh}, & \quad\forall s\in\mathcal{S},&\quad\forall h\in\mathcal{H} & \qquad\left(\underline{\rho}^u_{sh},\bar{\rho}^u_{sh}\right) \\
        \label{eq:opf-lp-ref-node-cnstr}& \theta^{ref}_{sh} = 0,                  & \quad\forall s\in\mathcal{S},&\quad\forall h\in\mathcal{H} & \qquad\left(\delta_{sh}\right).
    \end{alignat}
\end{subequations}

The objective function \eqref{eq:opf-lp-obj} is the cost observed by the system operator $C(\cdot)$ for the 24-hour period, while equation \eqref{eq:opf-lp-mkt-clrng-cnstr} enforces the power balance constraint in all nodes and equation \eqref{eq:opf-lp-pow-flow-cnstr} describes the relationship between power flows and voltage angles. Matrix $A\in M_{E\times N}\left(\R\right)$ in \eqref{eq:opf-lp-mkt-clrng-cnstr} is the incidence matrix of the electric power network and matrix $B\in M_{E\times E}\left(\R\right)$  in \eqref{eq:opf-lp-pow-flow-cnstr} is the diagonal matrix of the susceptances of the electric power network edges. Equations \eqref{eq:opf-lp-gen-lim-cnstrs}-\eqref{eq:opf-lp-unsd-lim-cnstrs} describe the upper and lower limits on power generation, power flows, voltage angle difference in all edges, and unserved demand respectively. Equation \eqref{eq:opf-lp-ref-node-cnstr} defines the reference node of the power system.
The system operator in our formulation minimizes the following operating cost of the power system:
\begin{align}\label{eq:cyberattacker-cost}
        C\left(g,u\right) =& \sum_{s\in\mathcal{S},h\in\mathcal{H} }\left(c^{g}_{sh}\right)^T g_{sh} 
        + \sum_{s\in\mathcal{S}, h\in\mathcal{H} }\left(c^u_{sh}\right)^T u_{sh}.    
    \end{align}

\paragraph{DC-OPF as a Complementarity Problem.} The DC-OPF comprises a convex (linear) objective function and convex feasible set, defined by a set of linear equality and inequality constraints, therefore \eqref{eq:opf-lp} is a convex optimization problem. Hence,
a point that satisfies the Karush Kuhn Tacker (KKT) conditions of \eqref{eq:opf-lp} is also a minimizer of \eqref{eq:opf-lp} \citep{boyd2004}. Moreover, we can recast the KKT conditions of a convex optimization problem as a Mixed Complementarity Problem (MCP) \cite{gabriel2012}. The MCP reformulation allows us to represent in a single problem the physical variables $(g,u,f,\theta)$ and prices and rents $(\pi^d, \pi^f, \bar{\rho}^g, \bar{\rho}^f)$ of problem \eqref{eq:opf-lp}. In addition, we will see in Section \ref{subsec:methodology-ul-cyber} that the MCP reformulation will facilitate the formulation and solution of the cyberattacker bilevel problem. The conditions \eqref{eq:opf-mcp} describe the MCP formulation of the KKT conditions of \eqref{eq:opf-lp}.

\begin{subequations}\label{eq:opf-mcp}
    \begin{alignat}{7}
        \label{eq:opf-mcp-kkt-gen}
        0 =&\quad c^{g,l}_{sh}
        - \underline{\rho}^g_{sh} + \bar{\rho}^g_{sh} - \pi^d_{sh} &\quad \bot \quad & g_{sh} \quad & \text{: free}, &\quad\forall s\in\mathcal{S}, &\quad \forall h\in\mathcal{H} \\
        \label{eq:opf-mcp-kkt-flow}
        0 =&\quad 
        A\pi^d_{sh}+\pi^f_{sh}-\underline{\rho}^{f}_{sh}+\bar{\rho}^f_{sh} &\quad \bot \quad & f_{sh} \quad & \text{: free}, &\quad\forall s\in\mathcal{S},&\quad \forall h\in\mathcal{H} \\
        \label{eq:opf-mcp-kkt-angle}
        0 =&\quad A^T\left(B\pi^f_{sh} - \underline{\rho}^\theta_{sh}+\bar{\rho}^\theta_{sh} \right)+e_{\text{ref}}\delta_{sh} &\quad \bot \quad & \theta_{sh} \quad & \text{: free}, &\quad\forall s\in\mathcal{S},&\quad \forall h\in\mathcal{H} \\
        \label{eq:opf-mcp-kkt-unsd-dem}
        0 =&\quad c^u_{sh} - \underline{\rho}^u_{sh} + \bar{\rho}^u_{sh} - \pi^d_{sh} &\quad \bot \quad & u_{sh} \quad & \text{: free}, &\quad\forall s\in\mathcal{S},&\quad \forall h\in\mathcal{H} \\
        \label{eq:opf-mcp-gen-lo}
        0 \leq &\quad g_{sh} - \underline{g}_{sh} &\quad \bot \quad &  \underline{\rho}^g_{sh} & \geq 0, &\quad\forall s\in\mathcal{S},&\quad \forall h\in\mathcal{H} \\    
        \label{eq:opf-mcp-gen-up}
        0 \leq &\quad \bar{g}_{sh} - g_{sh} &\quad \bot \quad & \bar{\rho}^g_{sh} & \geq 0, &\quad\forall s\in\mathcal{S},&\quad \forall h\in\mathcal{H} \\
        \label{eq:opf-mcp-flow-lo}
        0 \leq &\quad f_{sh} - \underline{f}_{sh} &\quad \bot \quad &  \underline{\rho}^f_{sh} & \geq 0, &\quad\forall s\in\mathcal{S},&\quad \forall h\in\mathcal{H} \\    
        \label{eq:opf-mcp-flow-up}
        0 \leq &\quad \bar{f}_{sh} - f_{sh} &\quad \bot \quad & \bar{\rho}^f_{sh} & \geq 0, &\quad\forall s\in\mathcal{S},&\quad \forall h\in\mathcal{H} \\
        \label{eq:opf-mcp-angle-lo}
        0 \leq &\quad A\theta_{sh} - \underline{\theta}_{sh} &\quad \bot \quad &  \underline{\rho}^\theta_{sh} & \geq 0, &\quad\forall s\in\mathcal{S},&\quad \forall h\in\mathcal{H} \\    
        \label{eq:opf-mcp-angle-up}
        0 \leq &\quad \bar{\theta}_{sh} - A\theta_{sh} &\quad \bot \quad & \bar{\rho}^\theta_{sh} & \geq 0, &\quad\forall s\in\mathcal{S},&\quad \forall h\in\mathcal{H} \\
        \label{eq:opf-mcp-unsd-lo}
        0 \leq &\quad u_{sh} &\quad \bot \quad &  \underline{\rho}^u_{sh} & \geq 0, &\quad\forall s\in\mathcal{S},&\quad \forall h\in\mathcal{H} \\    
        \label{eq:opf-mcp-unsd-up}
        0 \leq &\quad \left(d^w_{sh}+d^o_{sh}\right) - u_{sh} &\quad \bot \quad & \bar{\rho}^u_{sh} & \geq 0, &\quad\forall s\in\mathcal{S},&\quad \forall h\in\mathcal{H} \\
        \label{eq:opf-mcp-angle-ref}
        0 = &\quad \theta^{\text{ref}}_{sh} &\quad \bot \quad & \delta_{sh} & \text{: free}, &\quad\forall s\in\mathcal{S},&\quad \forall h\in\mathcal{H}. 
    \end{alignat}
\end{subequations} 

Following the literature on complementarity problems \citep{luo1996,ralph2007,siddiqui2013, sankaranarayanan2018}, we use ``$\bot$" to describe the complementary relationship between the left-hand and right-hand side conditions in each equation. The notation depicts that an equilibrium satisfies the equations in both sides of all complementarity conditions and at least one side is equal to zero in each condition. Following the MCP notation, we also denote variables without lower bounds as ``free". For ease of notation, we denote all optimal power flow variables by 

$$
    y=\left(g,f,u,\theta,\pi^d,\pi^f,\underline{\rho}^g,\bar{\rho}^g,\underline{\rho}^f,\bar{\rho}^f,\underline{\rho}^\theta,\bar{\rho}^\theta,\underline{\rho}^u_{sh},\bar{\rho}^u_{sh},\delta\right)\in\mathcal{Y}. 
$$

We can write \eqref{eq:opf-mcp} compactly as
\begin{subequations}\label{eq:opf-mcp-compact}
    \begin{alignat}{8}
    \label{eq:opf-mcp-compact-eq}&0=&\quad F_{1sh}(y_1,y_2) \quad&\bot&\quad y_{1sh} \quad&\text{: free}, &\quad \forall s\in\mathcal{S},&\quad h\in\mathcal{H} & \\
    \label{eq:opf-mcp-compact-ineq}&0 \leq&\quad F_{2sh}(y_1,y_2) \quad&\bot&\quad y_{2sh} \quad&\geq 0, &\quad \forall s\in\mathcal{S},&\quad h\in\mathcal{H}, &
    \end{alignat}
\end{subequations} 
where 
$$
y=\left(y_1,y_2\right)\in\mathcal{Y}_1\times\mathcal{Y}_2=\prod_{s\in\mathcal{S},h\in\mathcal{H}}\mathcal{Y}_{1sh}\times\prod_{s\in\mathcal{S},h\in\mathcal{H}}\mathcal{Y}_{2sh},
$$

$$
y_{1sh}=\left(g_{sh},u_{sh},f_{sh},\theta_{sh},\delta_{sh}\right)\in\mathcal{Y}_{1sh}=\prod_{i=1}^5\mathcal{Y}_{1sh}^{(i)},
$$
$$
y_{2sh}=\left(\underline{\rho}^g_{sh},\bar{\rho}^g_{sh},\underline{\rho}^f_{sh},\bar{\rho}^f_{sh},\underline{\rho}^\theta_{sh},\bar{\rho}^\theta_{sh},\underline{\rho}^u_{sh},\bar{\rho}^u_{sh}\right)\in\mathcal{Y}_{2sh}=\prod_{i=1}^8\mathcal{Y}_{2sh}^{(i)}.
$$

Notice that $F_1, F_2$ are linear functions. An equilibrium of \eqref{eq:opf-mcp} is a point $y^*=\left(y_1^*,y_2^*\right)$ that satisfies all complementarity conditions \eqref{eq:opf-mcp-kkt-gen}-\eqref{eq:opf-mcp-angle-ref}. Since the original problem is convex, then a point $\left(g^*,f^*,u^*,\theta^*\right)$ that satisfies the KKT conditions \eqref{eq:opf-mcp-compact} is also a minimizer of \eqref{eq:opf-lp} \cite{boyd2004}.

\subsection{Upper-Level Decisions:  Model of the Cyberattacker}\label{subsec:methodology-ul-cyber}

The \textit{Cyberattack} scenario designed in Section \ref{sec:scenario-design} assumes that the cyberattacker maximizes the operational disruption in electric power infrastructure by targeting SCADA devices and PLC interfaces. The cyberattacker strategy includes compromising power generation capacity $\left(z^g_{sh}\in\R^G\right)$, \textit{i.e.,} the upper limit, transmission limits $(z^f_{sh}\in\R^E)$, and limits on voltage angle differences $\left(z^\theta_{sh}\in\R^E\right)$. Hence, the physical power generation capacity, transmission, and limits on voltage angle differences under a cyberattack are modeled as

\begin{subequations}\label{eq:attacker-compromised-capacity}
    \begin{alignat}{6}
            \label{eq:attacker-compromised-capacity-gen}& \underline{g}_{sh}\quad& \leq&\quad g_{sh} &\leq&\quad \bar{g}_{sh} - z^g_{sh}, \quad&\quad \forall s\in\mathcal{S},& \quad\forall h\in\mathcal{H}& \\
            \label{eq:attacker-compromised-flow}& \underline{f}_{sh} - z^f_{sh}\quad& \leq&\quad f_{sh} &\leq&\quad \bar{f}_{sh}- z^f_{sh}, \quad&\quad \forall s\in\mathcal{S},& \quad\forall h\in\mathcal{H}&  \\
            \label{eq:attacker-compromised-capacity-angles}&\underline{\theta}_{sh} - z^\theta_{sh}\quad& \leq&\quad A\theta_{sh} \quad&\leq&\quad \bar{\theta}_{sh} - z^\theta_{sh}, &\quad \forall s\in\mathcal{S},& \quad\forall h\in\mathcal{H}. & 
    \end{alignat}
\end{subequations}

Given the ability of the cyberattacker to compromise components of electric power infrastructure as modeled in \eqref{eq:attacker-compromised-capacity-gen}-\eqref{eq:attacker-compromised-capacity-angles}, the bilevel diruption-maximizing problem of the cyberattacker can be formulated as

\begin{subequations}\label{eq:attacker-bilevel}
    \begin{alignat}{4}
        &\multispan3{%
            $\begin{array}{l}
             \max\limits_{\begin{array}{l}
            z^g_{sh}\in\R^G,z^f_{sh}\in\R^E, \\
            z^\theta_{sh}\in\R^E,y\in\mathcal{Y}
            \end{array}} 
            \Tilde{C}\left(y,z^g,z^f,z^\theta\right) 
            \end{array}$}\nonumber   \\
        \label{eq:attacker-bilevel-obj}\text{s.t.} & && \\
        \label{eq:attacker-bilevel-lowerlevel}& y\in\arg\min\limits_y\left\{C(y;z^g,z^f,z^\theta)\,\, |\,\, \eqref{eq:opf-lp-mkt-clrng-cnstr}-\eqref{eq:opf-lp-pow-flow-cnstr},\eqref{eq:attacker-compromised-capacity-gen} - \eqref{eq:attacker-compromised-capacity-angles}, \eqref{eq:opf-lp-unsd-lim-cnstrs}- \eqref{eq:opf-lp-ref-node-cnstr}\right\} & \\ 
        \label{eq:attacker-bilevel-zG_cap}& 0\leq z^g_{sh} \leq \bar{g}_{sh} , & \quad \forall s\in\mathcal{S},& \quad\forall h\in\mathcal{H} &  \\
        \label{eq:attacker-bilevel-zF_cap}& 0\leq z^f_{sh} \leq \bar{f}_{sh} , & \quad \forall s\in\mathcal{S},& \quad\forall h\in\mathcal{H} &  \\
        \label{eq:attacker-bilevel-zTHETA_cap}& 0\leq z^\theta_{sh} \leq \bar{\theta}_{sh} , & \quad \forall s\in\mathcal{S},& \quad\forall h\in\mathcal{H} &  \\
        \label{eq:attacker-bilevel-budget}&  \sum_{s\in\mathcal{S}}\left(\Tilde{c}^{g}_{sh}\right)^T z^g_{sh} 
              + \sum_{s\in\mathcal{S}}\left(\Tilde{c}^{f}_{sh}\right)^T z^f_{sh} + \sum_{s\in\mathcal{S}}\left(\Tilde{c}^\theta_{sh}\right)^T z^\theta_{sh} \leq \Tilde{b}_{s}, &\quad \forall s\in\mathcal{S}. &
    \end{alignat}
\end{subequations} 

The  objective function of the cyberattacker in \eqref{eq:attacker-bilevel-obj} is a measure of the target disruption $\Tilde{C(\cdot)}$, while \eqref{eq:attacker-bilevel-lowerlevel} enforces that all lower-level variables are solutions to the DC-OPF. Equations \eqref{eq:attacker-bilevel-zG_cap}-\eqref{eq:attacker-bilevel-zTHETA_cap} capture that compromised generation and transmission capacity, and limits on voltage angle differences can not exceed the respective physical capacities and limits. However, a cyberattacker can target intelligent electronic devices of substations, \textit{e.g.,} transmission line protective relays \citep{yew-meng2021}, and generators, \textit{e.g.,} manipulating the automatic generation control signals \citep{aditya2015}, to constrain power flows and generation. Equation \eqref{eq:attacker-bilevel-budget} describes the capabilities of the cyberattacker in each hour $h$ of each season $s$. Compromising generation and transmission capacity, and voltage angle limits requires committing resources relevant to the vulnerability of each component, \textit{e.g.,} work-hours, software and hardware infrastructure, or cost of employing specialized personnel, with corresponding costs $\Tilde{c}^g_{sh}, \Tilde{c}^f_{sh}, \Tilde{c}^\theta_{sh}$ with respect to the total available resources $\Tilde{b}_{s}>0$ in season $s$. In Section \ref{sec:scenario-design} we assume that in the default \textit{Cyberattack} scenario, the cost of attacking transmission $\left(\Tilde{c}^f_{sh}, \Tilde{c}^\theta_{sh}\right)$ is significantly higher than generation $\left(\Tilde{c}^g_{sh}\right)$. Given the difficulty of quantifying component vulnerabilities, we assign $\Tilde{c}^f_{she}=5c^g_{sht}$ for all $e\in\mathcal{E}$, $t\in\mathcal{T}$ in the default \textit{Cyberattack} scenario and conduct sensitivity analysis in Section \ref{sec:results}. In our formulation, the cyberattacker targets unserved electricity demand, \textit{i.e.,}
$$
    \Tilde{C}\left(y,z^g,z^f,z^\theta\right) = \sum_{s\in\mathcal{S},\in\mathcal{H}} \left(c^u_{sh}\right)^T u_{sh}.
$$

Notice that \eqref{eq:attacker-bilevel} is an optimization problem where \eqref{eq:attacker-bilevel-lowerlevel} yields an equilibrium of the electricity market \eqref{eq:opf-lp}. Therefore, \eqref{eq:attacker-bilevel} is a Mathematical Program with Equilibrium Constraints \cite{luo1996}. Finding a global optimum of an MPEC is a combinatorial problem \cite{luo1996}, thus computationally demanding. In the next section we describe how to efficiently solve large-scale instances of \eqref{eq:attacker-bilevel}.

\subsection{Bilevel Optimization of the Cyberattacker: Solution Algorithm}\label{subsec:methodology-bilevel-algorithm}

In Section \ref{subsec:methodology-ll-dcopf} we showed that finding a minimizer of the DC-OPF formulation  \eqref{eq:opf-lp} is equivalent to finding a solution $y^*$ of the MCP \eqref{eq:opf-mcp-compact}. Therefore, we can substitute \eqref{eq:attacker-bilevel-lowerlevel} with the respective complementarity conditions \eqref{eq:opf-mcp-compact} and problem \eqref{eq:attacker-bilevel} becomes a Mathematical Program with Complementarity Constraints (MPCC), a special case of an MPEC \cite{ralph2007}, which allows us to model in a single framework the physical and technological (primal), and economic (dual) variables of the lower-level problem \eqref{eq:attacker-bilevel-lowerlevel}. Problem \eqref{eq:attacker-mpcc} is the MPCC reformulation of the bilevel optimizaiton of the cyberattacker in \eqref{eq:attacker-bilevel}.

\begin{subequations}\label{eq:attacker-mpcc}
    \begin{alignat}{4}
        \text{MPCC}:=&\multispan3{%
        $\begin{array}{ll}
             \max\limits_{\begin{array}{l}
            z^g_{sh}\in\R^G,z^f_{sh}\in\R^E, \\
         z^\theta_{sh}\in\R^E,y_1\in\mathcal{Y}_1,y_2\in\mathcal{Y}_2
         \end{array}} 
         \sum\limits_{s\in\mathcal{S},\in\mathcal{H}} \left(c^u_{sh}\right)^T u_{sh} 
        \end{array}$}\nonumber   \\
        \label{eq:attacker-mpcc-obj}\text{s.t.} & && \\
        \label{eq:attacker-mpcc-lowerlevel-equality}& F_{1sh}(y_1,y_2;z^g,z^f,z^\theta)  = 0 & \quad \forall s\in\mathcal{S},& \quad \forall h\in\mathcal{H}& \\ 
        \label{eq:attacker-mpcc-lowerlevel-complementarity}& 0 \leq  y_{2sh} \quad \bot\quad  F_{2sh}(y_1,y_2;z^g,z^f,z^\theta)  \geq 0 &\quad \forall s\in\mathcal{S},& \quad \forall h\in\mathcal{H}& \\
        \label{eq:attacker-mpcc-zG_cap}& 0\leq z^g_{sh} \leq \bar{g}_{sh} , & \quad \forall s\in\mathcal{S},& \quad \forall h\in\mathcal{H}&  \\
        \label{eq:attacker-mpcc-zF_cap}& 0\leq z^f_{sh} \leq \bar{f}_{sh} , & \quad \forall s\in\mathcal{S},& \quad \forall h\in\mathcal{H}&  \\
        \label{eq:attacker-mpcc-zTHETA_cap}& 0\leq z^\theta_{sh} \leq \bar{\theta}_{sh} , & \quad \forall s\in\mathcal{S},& \quad \forall h\in\mathcal{H}&  \\
        \label{eq:attacker-mpcc-budget}&  \sum_{s\in\mathcal{S}}\left(\Tilde{c}^{g}_{sh}\right)^T z^g_{sh} 
              + \sum_{s\in\mathcal{S}}\left(\Tilde{c}^{f}_{sh}\right)^T z^f_{sh} + \sum_{s\in\mathcal{S}}\left(\Tilde{c}^\theta_{sh}\right)^T z^\theta_{sh} \leq \Tilde{b}_{s}, &\quad \forall s\in\mathcal{S}&
    \end{alignat}
\end{subequations} 

Although all functions in \eqref{eq:attacker-mpcc} are linear, the MPCC is a nonlinear optimization problem since the complementarity conditions \eqref{eq:attacker-mpcc-lowerlevel-complementarity} imply that either $y_{2,i}$ or $F_{2,i}(y_1,y_2)$ are zero for all $i\in|\mathcal{Y}_2|$. Finding a global optimum of a nonlinear optimization problem can be computationally challenging, and many commercial solvers compute a local optimum \citep{scheel2000}. To compute a global optimum of \eqref{eq:attacker-mpcc}, we can recast the complementarity conditions using binary variables, and reformulate the MPCC into a Mixed Integer Linear Program (MILP) \cite{luo1996}. The MILP reformulation requires introducing as many binary variables as complementarity constraints in \eqref{eq:opf-mcp}, which in our problem are in the order of tens of thousands\footnote{The complementarity conditions include roughly $9 \text{ (generation technologies)}\times16 \text{ (nodes)}\times24 \text{ (hours)}\times3 \text{ (variables } g_{sh},\underline{\rho}^g_{sh},\bar{\rho}^g_{sh})\approx10,000$ generation-related conditions and variables in \eqref{eq:opf-mcp-kkt-gen},\eqref{eq:opf-mcp-gen-lo}-\eqref{eq:opf-mcp-gen-up}; $17 \text{ (edges)}\times24 \text{ (hours)}\times3 \text{ (variables } f_{sh},\underline{\rho}^f_{sh},\bar{\rho}^f_{sh})\approx1,000$ flow-related conditions and variables in \eqref{eq:opf-mcp-kkt-flow},\eqref{eq:opf-mcp-flow-lo}-\eqref{eq:opf-mcp-flow-up}; and $16 \text{ (nodes)}\times24 \text{ (hours)}\times 6 \text{ (variables } \theta_{sh}, u_{sh}, \underline{\rho}^\theta_{sh},\bar{\rho}^\theta_{sh}, \underline{\rho}^u_{sh},\bar{\rho}^u_{sh}, \delta_{sh})\approx2,000$ node-related conditions and variables in \eqref{eq:opf-mcp-kkt-angle}-\eqref{eq:opf-mcp-kkt-unsd-dem}, \eqref{eq:opf-mcp-angle-lo}-\eqref{eq:opf-mcp-angle-ref}.}. For the MILP reformulation, we introduce the following binary variables and parameters\footnote{Note that $y_{2sh}=\left(\underline{\rho}^g_{sh},\bar{\rho}^g_{sh},\underline{\rho}^f_{sh},\bar{\rho}^f_{sh},\underline{\rho}^\theta_{sh},\bar{\rho}^\theta_{sh},\underline{\rho}^u_{sh},\bar{\rho}^u_{sh}\right)\in\prod_{i=1}^8\R^{\kappa_i}_{\geq 0}$.} $$
\gamma_{sh}\in\prod_{i=1}^8\left\{0,1\right\}^{\kappa_{i}}, \quad M_{sh}\in\prod_{i=1}^8\R_{\geq 0}^{\kappa_{i}}, \quad \kappa_i=\left\lvert \mathcal{Y}^{(i)}_{2sh}\right\rvert, \qquad \forall s\in\mathcal{S}, \,\, h\in\mathcal{H}.
$$

The MILP reformulation of the MPCC in \eqref{eq:attacker-mpcc} is

\begin{subequations}\label{eq:attacker-milp}
    \begin{alignat}{4}
        \text{MPCC-MILP:=}&\multispan3{%
        $\begin{array}{ll}
             \max\limits_{\begin{array}{l}
            z^g_{sh}\in\R^G,z^f_{sh}\in\R^E, \\
         z^\theta_{sh}\in\R^E,y_1\in\mathcal{Y}_1,y_2\in\mathcal{Y}_2
         \end{array}} 
         \sum\limits_{s\in\mathcal{S},h\in\mathcal{H}} \left(c^u_{sh}\right)^T u_{sh}
        \end{array}$}\nonumber   \\
        \label{eq:attacker-milp-obj}\text{s.t.} & && \\
        \label{eq:attacker-milp-lowerlevel-equality}& F_{1sh}(y_1,y_2;z^g,z^f,z^\theta)  = 0, &\quad \forall s\in\mathcal{S},& \quad \forall h\in\mathcal{H}& \\ 
        \label{eq:attacker-milp-lowerlevel-binary-funct}& 0 \leq  F_{2sh}(y_1,y_2;z^g,z^f,z^\theta) \leq \left(\mathbf{1}_{sh}-\gamma_{sh}\right)M_{sh}, &\quad \forall s\in\mathcal{S},& \quad \forall h\in\mathcal{H}&\\
        \label{eq:attacker-milp-lowerlevel-binary-var}& 0 \leq  y_{2sh} \leq \gamma_{sh} \cdot M_{sh}, &\quad \forall s\in\mathcal{S},& \quad \forall h\in\mathcal{H}& \\
%
        \label{eq:attacker-milp-zG_cap}& 0\leq z^g_{sh} \,\,\leq \bar{g}_{sh} , & \quad \forall s\in\mathcal{S},& \quad \forall h\in\mathcal{H}&  \\
        \label{eq:attacker-milp-zF_cap}& 0\leq z^f_{sh} \,\,\leq \bar{f}_{sh} , & \quad \forall s\in\mathcal{S}& \quad \forall h\in\mathcal{H}&  \\
        \label{eq:attacker-milp-zTHETA_cap}& 0\leq z^\theta_{sh} \,\,\leq \bar{\theta}_{sh} , & \quad \forall s\in\mathcal{S},& \quad \forall h\in\mathcal{H}&  \\
        \label{eq:attacker-milp-budget}&  \sum_{h\in\mathcal{H}}\left(\Tilde{c}^{g}_{sh}\right)^T z^g_{sh} 
        + \sum_{h\in\mathcal{H}}\left(\Tilde{c}^{f}_{sh}\right)^T z^f_{sh} + \sum_{h\in\mathcal{H}}\left(\Tilde{c}^\theta_{sh}\right)^T z^\theta_{sh} \leq \Tilde{b}_{s}, &\quad \forall s\in\mathcal{S}& &
    \end{alignat}
\end{subequations} 
where $\mathbf{1}_{sh}\in\mathcal{Y}_{2sh}$ is a vector of all ones and ``$\cdot$" in equations \eqref{eq:attacker-milp-lowerlevel-binary-funct}-\eqref{eq:attacker-milp-lowerlevel-binary-var} is the element-wise product between the two vectors of each bilinear term. Equations \eqref{eq:attacker-milp-lowerlevel-binary-var}, \eqref{eq:attacker-milp-lowerlevel-binary-funct} are equivalent to \eqref{eq:attacker-mpcc-lowerlevel-complementarity}\footnote{When $\gamma_{sh,i}=1$ then $F_{2sh,i}=0$ and $0\leq y_{2sh,i} \leq M_{sh,i}$. Similarly, when $\gamma_{sh,i}=0$ then $y_{2sh,i}=0$ and $0 \leq F_{2sh,i} \leq M_{sh,i}$. Thus, equations \eqref{eq:attacker-milp-lowerlevel-binary-var}, \eqref{eq:attacker-milp-lowerlevel-binary-funct} enforce that $y_{2sh,i}, F_{2sh,i}(y_1,y_2)$ are non-negative and at least one is zero.}. If $\left(z^{g*},z^{f*},z^{\theta*},y_1^*,y_2^*\right)$ is a global optimum of \eqref{eq:attacker-mpcc} and parameter $M_i>\lvert\lvert\left(z^{g*},z^{f*},z^{\theta*},y_1^*,y_2^*\right)\rvert\rvert_{\infty}$, then a solution of the MILP in \eqref{eq:attacker-milp} is a global optimum of the original MPCC \eqref{eq:attacker-mpcc} \cite{feng2018}.


Finding a global optimum of the MPCC \eqref{eq:attacker-mpcc} is a combinatorial problem \cite{luo1996} and requires significant computational power. In our formulation, the number of binary variables of the MPCC-MILP \eqref{eq:attacker-milp} is in the order of tens of thousands, which exacerbates the problem of available computational resources and long computation time. The Nonlinear (NLP) reformulation of a MPCC addresses the problem of computational power and is the most widely used alternative to the MILP reformulation for large-scale MPCCs \cite{leyffer2006,leyffer2010}. Under the NLP reformulation, we cast the complementarity conditions \eqref{eq:attacker-mpcc-lowerlevel-complementarity} as equality and inequality constraints \eqref{eq:attacker-nlp-lowerlevel-bl2}-\eqref{eq:attacker-nlp-lowerlevel-y2} to derive the Nonlinear Programming (NLP) reformulation of the MPCC \eqref{eq:attacker-mpcc}, \textit{i.e.,}

\begin{subequations}\label{eq:attacker-nlp}
    \begin{alignat}{4}
        \text{MPCC-NLP}:=&\multispan3{%
        $\begin{array}{ll}
             \max\limits_{\begin{array}{l}
            z^g_{sh}\in\R^G,z^f_{sh}\in\R^E, \\
         z^\theta_{sh}\in\R^E,y_1\in\mathcal{Y}_1,y_2\in\mathcal{Y}_2
         \end{array}} 
         \sum\limits_{s\in\mathcal{S},\in\mathcal{H}} \left(c^u_{sh}\right)^T u_{sh} 
        \end{array}$}\nonumber   \\
        \label{eq:attacker-nlp-obj}\text{s.t.} & && \\
        \label{eq:attacker-nlp-lowerlevel-equality}& F_{1sh}(y_1,y_2;z^g,z^f,z^\theta) = 0 & \quad \forall s\in\mathcal{S},& \quad \forall h\in\mathcal{H}& \\ 
        \label{eq:attacker-nlp-lowerlevel-bl2}&    F_{2sh}(y_1,y_2;z^g,z^f,z^\theta)\cdot y_{2sh} = 0 &\quad \forall s\in\mathcal{S},& \quad \forall h\in\mathcal{H}& \\
        \label{eq:attacker-nlp-lowerlevel-F2}& 0 \leq  F_{2sh}(y_1,y_2;z^g,z^f,z^\theta)  &\quad \forall s\in\mathcal{S},& \quad \forall h\in\mathcal{H}& \\
        \label{eq:attacker-nlp-lowerlevel-y2}& 0 \leq  y_{2sh} &\quad \forall s\in\mathcal{S},& \quad \forall h\in\mathcal{H}& \\
        \label{eq:attacker-nlp-zG_cap}& 0\leq z^g_{sh} \leq \bar{g}_{sh} , & \quad \forall s\in\mathcal{S},& \quad \forall h\in\mathcal{H}&  \\
        \label{eq:attacker-nlp-zF_cap}& 0\leq z^f_{sh} \leq \bar{f}_{sh} , & \quad \forall s\in\mathcal{S},& \quad \forall h\in\mathcal{H}&  \\
        \label{eq:attacker-nlp-zTHETA_cap}& 0\leq z^\theta_{sh} \leq \bar{\theta}_{sh} , & \quad \forall s\in\mathcal{S},& \quad \forall h\in\mathcal{H}&  \\
        \label{eq:attacker-nlp-budget}&  \sum_{h\in\mathcal{H}}\left(\Tilde{c}^{g}_{sh}\right)^T z^g_{sh} 
              + \sum_{h\in\mathcal{H}}\left(\Tilde{c}^{f}_{sh}\right)^T z^f_{sh} + \sum_{h\in\mathcal{H}}\left(\Tilde{c}^\theta_{sh}\right)^T z^\theta_{sh} \leq \Tilde{b}_{s}, &\quad \forall s\in\mathcal{S}&
    \end{alignat}
\end{subequations}

Using a NLP solver to solve the MPCC-NLP is computationally more efficient for many MPCC problems \cite{siddiqui2013,feng2018}. However, the bilinear term \eqref{eq:attacker-nlp-lowerlevel-bl2} implies that conventional NLP solvers can converge to a stationary point or local optimum of the
MPCC-NLP \eqref{eq:attacker-nlp} \cite{leyffer2010}. Moreover, MPCCs, like MPECs, can have disjoint feasible sets \cite{huppmann2018}, hence the solver optimum depends on the starting point. In our MPCC, there exists a tradeoff between finding a global optimum using the MILP reformulation and computational efficiency using the NLP reformulation. The first is more computationally demanding while the second may converge to a local optimum. To overcome these challenges, we will use a heuristic to approximate the solution of the MILP. Consequently, we will initiate the MPCC-NLP \eqref{eq:attacker-nlp} at a point informed by the structure of our problem. Specifically, for each season $s$ and hour $h$ in our formulation, functions $F_{1sh},F_{2sh}$ depend only on $y_{1sh}, y_{2sh}$. Hence, only constraint \eqref{eq:attacker-nlp-budget} couples decisions across time periods, while all remaining constraints \eqref{eq:attacker-nlp-lowerlevel-equality}-\eqref{eq:attacker-nlp-zTHETA_cap} and all variables are decoupled in $s,h$. Moreover, the objective function is additive. Therefore, we can approximate the MPCC-MILP reformulation in \eqref{eq:attacker-milp} by maximizing a series of hourly problems $(c^u_{sh})^T u_{sh}$ for each $s$ and $h$ subject to an hourly budget

$$
    \Tilde{b}_{sh} = \frac{\Tilde{b}_s}{H}, \qquad \forall s \in \mathcal{S}, \qquad h\in\mathcal{H}.
$$

The decoupled MPCC-MILP(s,h) for each season $s$ and hour $h$ is

\begin{subequations}\label{eq:attacker-dmilp}
    \begin{alignat}{4}
        \text{DMPCC-MILP(s,h):=}&\multispan3{%
        $\begin{array}{ll}
             \max\limits_{\begin{array}{l}
            z^g_{sh}\in\R^G,z^f_{sh}\in\R^E, \\
         z^\theta_{sh}\in\R^E,y_1\in\mathcal{Y}_1,y_2\in\mathcal{Y}_2
         \end{array}} 
         \sum\limits_{s\in\mathcal{S},h\in\mathcal{H}} \left(c^u_{sh}\right)^T u_{sh}
        \end{array}$}\nonumber   \\
        \label{eq:attacker-dmilp-obj}\text{s.t.} & && \\
        \label{eq:attacker-dmilp-lowerlevel-equality}& F_{1sh}(y_1,y_2;z^g,z^f,z^\theta)  = 0, &\quad \forall s\in\mathcal{S},& \quad \forall h\in\mathcal{H}& \\ 
        \label{eq:attacker-dmilp-lowerlevel-binary-funct}& 0 \leq  F_{2sh}(y_1,y_2;z^g,z^f,z^\theta) \leq \left(\mathbf{1}_{sh}-\gamma_{sh}\right)M_{sh}, &\quad \forall s\in\mathcal{S},& \quad \forall h\in\mathcal{H}&\\
        \label{eq:attacker-dmilp-lowerlevel-binary-var}& 0 \leq  y_{2sh} \leq \gamma_{sh} \cdot M_{sh}, &\quad \forall s\in\mathcal{S},& \quad \forall h\in\mathcal{H}& \\
        \label{eq:attacker-dmilp-zG_cap}& 0\leq z^g_{sh} \,\,\leq \bar{g}_{sh} , & \quad \forall s\in\mathcal{S},& \quad \forall h\in\mathcal{H}&  \\
        \label{eq:attacker-dmilp-zF_cap}& 0\leq z^f_{sh} \,\,\leq \bar{f}_{sh} , & \quad \forall s\in\mathcal{S}& \quad \forall h\in\mathcal{H}&  \\
        \label{eq:attacker-dmilp-zTHETA_cap}& 0\leq z^\theta_{sh} \,\,\leq \bar{\theta}_{sh} , & \quad \forall s\in\mathcal{S},& \quad \forall h\in\mathcal{H}&  \\
        \label{eq:attacker-dmilp-budget}&  \left(\Tilde{c}^{g}_{sh}\right)^T z^g_{sh} 
        + \left(\Tilde{c}^{f}_{sh}\right)^T z^f_{sh} + \left(\Tilde{c}^\theta_{sh}\right)^T z^\theta_{sh} \leq \Tilde{b}_{sh}, &\quad \forall s\in\mathcal{S},& \quad \forall h\in\mathcal{H}&
    \end{alignat}
\end{subequations} 

The solution $\left(z^{g(1)}_{sh},z^{f(1)}_{sh},z^{\theta(1)}_{sh},y^{(1)}_{1sh},y^{(1)}_{2sh}\right)$ of the DMPCC-MILP \eqref{eq:attacker-dmilp} may not optimize the MPCC-MILP \eqref{eq:attacker-milp}, since the decoupled formulation assumes the attacker's budget allocation to be homogeneous across time periods. We then initiate the MPCC-NLP with $\left(z^{g(1)}_{sh},z^{f(1)}_{sh},z^{\theta(1)}_{sh},y^{(1)}_{1sh},y^{(1)}_{2sh}\right)$ to compute the cyberattacker's actions under a more disruptive budget allocation. Table \ref{table:milp-algorithm} describes the steps of the proposed solution algorithm.

\begin{table}[h!]
    \begin{tabular}{l l l l}
        \hline 
        \multispan3{\textbf{Algorithm.}} \\ 
        \hline
        \hline
        \textbf{Begin}& & & \\
        &\textbf{For} $s\in\mathcal{S}$   &  \\
        & & \textbf{For} &$h\in\mathcal{H}$   \\
        & &    & \textbf{Find} $\left(z^{g(1)}_{sh},z^{f(1)}_{sh},z^{\theta(1)}_{sh},y^{(1)}_{1sh},y^{(1)}_{2sh}\right)$ solution of: \\
        & &   &DMPCC-MILP(s,h):=$\max\limits_{\begin{array}{l}
        z^{g}_{sh},z^{f}_{sh},z^{\theta}_{sh}, \\y_{1sh},y_{2sh}
        \end{array} 
        } \left(c^u_{sh}\right)^T u_{sh}$ \\
        & & & \qquad\qquad\qquad\qquad\quad\qquad\quad s.t. \eqref{eq:attacker-milp-lowerlevel-binary-funct}-\eqref{eq:attacker-milp-zTHETA_cap} \\
        & & & \qquad\qquad\qquad\qquad\qquad\quad\qquad\,\,\, $\left(\Tilde{c}^{g}_{sh}\right)^T z^g_{sh} 
        + \left(\Tilde{c}^{f}_{sh}\right)^T z^f_{sh} + \left(\Tilde{c}^\theta_{sh}\right)^T z^\theta_{sh} \leq \Tilde{b}_{sh}$ \\
        & & \textbf{End} & \\
        & \quad\textbf{End} & & \\
        &\multispan3{%
        \textbf{Initialize} MPCC-NLP with the decoupled DMPCC-MILP(s,h) solutions:} \\
        & &\multispan2{%
        $\left(z^{g},z^{f},z^{\theta},y_{1},y_{2}\right)^{(0)}= \prod\limits_{s\in\mathcal{S},h\in\mathcal{H}} \left(z^{g(1)}_{sh},z^{f(1)}_{sh},z^{\theta(1)}_{sh},y_{1}^{(1)},y_{2}^{(1)}\right)
        $.} \\
        & \multispan3{%
        \textbf{Find} $\left(z^{g*},z^{f*},z^{\theta*},y^{*}_{1},y^{*}_{2}\right)$ solution of:} \\
        & &\multispan2{%
        MPCC-NLP:=$\max\limits_{\begin{array}{l}
        z^{g}_{sh},z^{f}_{sh},z^{\theta}_{sh}, \\y_{1sh},y_{2sh}
        \end{array} 
        } \left(c^u_{sh}\right)^T u_{sh}$} \\
        & & & \qquad\qquad\quad s.t. \eqref{eq:attacker-nlp-lowerlevel-equality}-\eqref{eq:attacker-nlp-zTHETA_cap}. \\
        & & & \qquad \quad\qquad\quad\,\, $\sum\limits_{h\in\mathcal{H}}\left(\Tilde{c}^{g}_{sh}\right)^T z^g_{sh} 
        + \sum\limits_{h\in\mathcal{H}}\left(\Tilde{c}^{f}_{sh}\right)^T z^f_{sh} + \sum\limits_{h\in\mathcal{H}}\left(\Tilde{c}^\theta_{sh}\right)^T z^\theta_{sh} \leq \Tilde{b}_{s}$ \\
        \textbf{End}& & & \\
        \hline
    \end{tabular}
    \caption{ Decomposition algorithm of MILP reformulation \eqref{eq:attacker-milp} of MPCC \eqref{eq:attacker-mpcc}. Since the objective function of the original problem is additive and all constraints are disjoint in $s,h$ for fixed budget allocation, then we can retrieve an approximate solution to the MPCC-MILP under a homogeneous budget allocation by solving the DMPCC-MILP(s,h) for all $s\in\mathcal{S},h\in\mathcal{H}$. We then feed the DMPCC-MILP solution as a starting point to the MPCC-NLP and allow for flexible budget allocation between time periods to compute an optimal disruption also across time periods.}
    \label{table:milp-algorithm}
\end{table}

\subsection{Economy-Wide Model: Computable General Equilibrium}\label{subsec:windc}
We evaluate the ripple effects of disruptions in electric power networks across all economic sectors using the WiNDC model, developed and maintained by the University of Wisconsin, Madison \citep{rutherford2019}. The economy-wide cost of blackouts can be hard to derive. For example, in the 2003 Northeastern Blackout, the U.S. Department of Energy estimated losses between \$4 -- \$10 Billion in the U.S. \citep{usdoe2004} following the event. However, finding the Value of Lost Load (VOLL) under a power outage can be challenging and estimates vary. For example, VOLL is different between industrial, commercial and residential consumers and the estimation can be based on econometric modeling under multiple assumptions \citep{sullivan2009}. System operators employ a variety of methods to compute VOLL, including an \textit{ex-post} macroeconomic analysis of the contribution of an event to the reduction of the Gross Domestic Product \citep{ercot2013}. Despite the shortcomings of computing the VOLL using \textit{ex-post} macroeconomic analysis, the Electric Reliability Council of Texas, Inc. reports that the results can serve as a \textit{``reference point"}. Instead, 
we aim to provide an \textit{ex-ante} estimation of how transactions between sectors deviate from the \textit{Baseline} using a Computable General Equilibrium (CGE) model to improve the assessment of outages' riple effects on the economy.

The WiNDC model is a multi-regional, multi-sectoral CGE Model comprising 51 regions, including the 50 States and the District of Columbia denoted with their two-letter code, and 71 economic sectors. In New York State, outside the \textit{``Utilities"} sector, the \textit{``Residential"} sector consumes the most output of the ``Utilities" sector in New York State, followed by the \textit{``Construction"}, \textit{``State and Local Government Enterprises (SLGE)"}, and \textit{``Pipeline Transportation"} sectors. The output of each sector requires intermediate inputs from other sectors, labor and capital, and can cover regional demand and the demand for exports to other regions and the rest of the world. Domestic prices drive the optimal output and demand for production factors of each sector, while changes in regional prices drive inter-regional and international trade in each sector. Household consumption in WiNDC is driven by consumer preferences and regional price changes allowing households to substitute between goods in response to price changes. The model also allows for taxes and subsidies on intermediate demand, and import tariffs. Since 2019, the WiNDC model has been deployed to study economy-wide impacts to U.S. states of international food trade policies, climate policy design, and the impact of environmental regulation on regional labor markets. For climate policy design, Rutherford \textit{et al.,} \citep{rutherford2019} expand the core WiNDC module to account for energy intensity of sectoral supply chains to understand carbon leakage due to energy consumption at the subnational level, across U.S. states, while Becker \textit{et al.,} \citep{becker2022} further enhance WiNDC to understand the impact of climate policies on state-level labor supply across demographic groups. In the food sector, Ballistreri \textit{et al.,} \citep{balistreri2020}, employ WiNDC to understand adverse impacts of The
Phase One Trade agreement between
the United States and China across U.S. states \citep{ustr2020}. 

Figure \ref{fig:methodology} describes the coupling of the bilevel optimization problem with the CGE in our framework. First, the bilevel optimization problem assesses the availability of electricity supply of the New York State electric power network under a \textit{Heatwave}, \textit{Cyberattack}, or \textit{Compound Cyber-Physical Threat}. Then, WiNDC receives regional electricity supply as input and assesses the response of all other sectors and potential economic losses to the New York State economy. The implementation of the scenarios in Section~\ref{sec:scenario-design} in WiNDC required diverging from the core model in \citep{rutherford2019}. We adjust the base year of WiNDC from 2014 to 2018 to ensure that trade across sectors in the CGE are consistent with the operating year of the NYISO electric power network. For that, we introduce an exogenous total factor productivity, $tfp_{r}$ for each region $r$, which is set to the Gross Domestic Product (GDP) growth rate of each state \citep{bea2022} in all quantities, \textit{e.g.,} $X_{r,g}:=tfp_r\cdot X_{r,g}$ in equations (52)-(54) in \citep{rutherford2019} which become

\begin{alignat}{1}
    \label{eq:windc-supply}&\sum_s Y_{r,s} \bar{ys}_{r,s,g}+\bar{yh}_{r,g} - tfp_r X_{r,g} \bar{s}_{r,g} = 0 \\
    \label{eq:windc-distribution}&tfp_r X_{r,g} S^D_{r,g} - A_{r,g} D^D_{r,g} - \sum_m MS_{r,m} \bar{dm}_{r,g,m} = 0 \\
    \label{eq:windc-balance}&\sum_r \left( tfp_r X_{r,g} S^N_{r,g} - A_{r,g} D^N_{r,g} - \sum_m MS_{r,m} \bar{nm}_{r,g,m} \right) = 0
\end{alignat}

In WiNDC, set indices $r,s,g,m$ denote the WiNDC regions, sectors, goods produced by sectors, and margin adjustments for trade and transport. Equation \eqref{eq:windc-supply} regulates state-level supply, where $\bar{yh}_{r,g}$ denotes fixed household production, $\bar{ys}_{r,s,g}$ denotes baseline sectoral production in region $r$, $Y_{r,s}$  is the change of sectoral production compared to baseline, $\bar{s}_{r,g}$ is baseline regional supply, and $X_{r,g}$ is the change of regional supply compared to baseline. Equation \eqref{eq:windc-distribution} regulates allocation of supply to the state-level market, where $\bar{dm}_{g,m}$ denotes baseline state-level margin supply, $MS_{r,m}$ denotes change of  state-level margin supply compared to baseline, $D^D_{r,g}$ denotes state-level goods demand, $A_{r,g}$ is the regional allocation of goods, and $S^D_{r,g}$ denotes state-level supply. Equation \eqref{eq:windc-balance} ensures the balance of quantities at the national scale, where $S^N_{r,g}$ denotes national supply, $D^N_{r,g}$ denotes national goods demand, and $\bar{nm}_{r,g,m}$ denotes the baseline national supply of margin sectors.  

The New York State annual GDP in our model grows from \$1.47 Trillion in 2014 to \$1.57 Trillion in 2018, which translates to a daily GDP of \$4.31 Billion. The \textit{Compound Cyber-Physical Threat} causes a shortage of regional electricity supply, which leads to unserved demand. We implement the \textit{Compound Cyber-Physical Threat} scenarios in WiNDC by decreasing regional electricity supply provided by the ``Utilities" sector, similarly to the simulation of power outages impacts using a CGE in Rose \textit{et al.,} \citep{rose2004}. The ``Utilities" sector in this version of WiNDC comprises electricity, natural gas, and water/wastewater services. For that, we assume that a supply disruption in any of the three subsectors translates to an equivalent supply disruption to the services of the ``Utilities" sector as a whole. The  percentage decrease of regional electricity supply (variable $X_{r,g}$ for state $r$ and sector $g$) is equal to the percent of unserved load with respect to \textit{Baseline} load in the NYISO  model. Figure \ref{fig:methodology} shows the coupling between the bilevel optimization problem and the CGE which allows us to endogenously consider the operational disruptions within the electric power network under each scenario when assessing the response of all other economic sectors.

The computation of CGE parameters assumes yearly or 5-year steps. Therefore, a CGE simulates the long-term substitution between intermediate inputs and production factors for economic sectors, and between goods for households. Calculating robust substitution elasticities under extreme events requires often unavailable time-series data on the response of different sectors during the extreme event. For that, substitution elasticities under infrastructure disruptions are unavailable for most sectors, \textit{e.g.,} the food sector \citep{moynihan2022}. Nevertheless, conducting economy-wide analysis of a 24-hour disruption using a CGE can provide a lower bound on ripple effects and economic losses. Since sectors and consumers can substitute inputs and consumed goods easier in a CGE framework, we expect that the economic losses will be greater under more rigid substitution regimes and New York State would face \textit{at least} losses reported from the CGE.

\section{Results}\label{sec:results}
\subsection{Vulnerability of Regional Infrastructure Components}\label{subsec:results-default}

\begin{figure}[h!]
	    \centering	\includegraphics[width=1.0\textwidth]{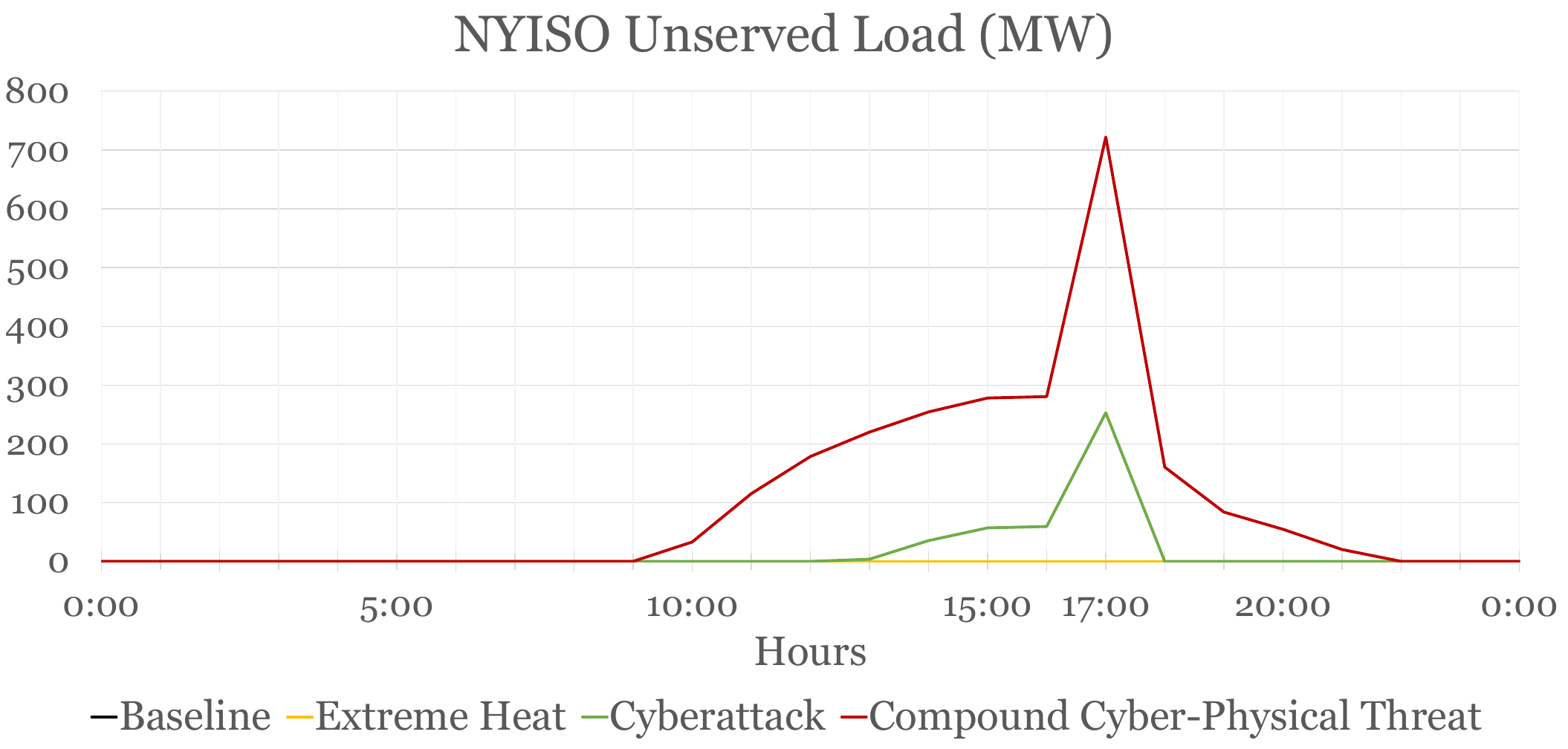}
		\caption{\label{fig:results-unservedload}\footnotesize NYISO Unserved Load (MW). Under the \textit{Heatwave} there is no unserved load. During peak load at 17:00 and under normal operating conditions, the \textit{Cyberattack}  compromises 253 MW (1\%) of NYISO load. For the same time period, the \textit{Compound Cyber-Physical Threat} causes load curtailment of 722MW, which is three times higher compared to the \textit{Cyberattack}.}
	\end{figure}

In this section, we identify vulnerable components of the NYISO grid and compare the magnitude and regional distribution of unserved load under the \textit{Heatwave}, \textit{Cyberattack}, and \textit{Compound Cyber-Physical Threat} scenarios as described in Section~\ref{sec:scenario-design}. \textbf{Figure \ref{fig:results-unservedload}} shows that the NYISO grid has the generation and transmission capacity to withstand the \textit{Heatwave} without shedding load, while the \textit{Cyberattack} under normal operating conditions results in 408 MW of unserved load in a 24-hour period, which is 0.1\% of total NYISO load or approximately 19,800 customers\footnote{We estimate number of customers by multiplying the percentage of unserved load with the total number of NYISO customers, as reported in the \textit{Baseline} scenario}. However, timing the \textit{Cyberattack} with the \textit{Heatwave} can compound the impact of the \textit{Cyberattack} by more than 3 times. Therefore, the \textit{Compound Cyber-Physical Threat} results in 2401 MWh of unserved load in a 24-hour period, which is almost 0.4\% of total NYISO load and affects almost 79,200 customers. However, the impact varies temporally, with the largest shedding  at 17:00, that is during the peak load hour. For comparison, under the \textit{Cyberattack}, peak load shedding reaches 253 MW, while under the \textit{Compound Cyber-Physical Threat}, peak load shedding reaches 722 MW.  

Timing the \textit{Cyberattack} excacerbates also the duration of load shedding. In the \textit{Cyberattack} scenario, load shedding happens for five hours, between 13:00-18:00, while in the \textit{Compound Cyber-Physical Threat} scenario, load shedding happens for 11 hours, between 10:00-21:00. Therefore, the \textit{Compound Cyber-Physical Threat} exacerbates both the impact and the duration of the \textit{Cyberattack}.

    \begin{figure}[h!]
	    \centering	\includegraphics[width=1.0\textwidth]{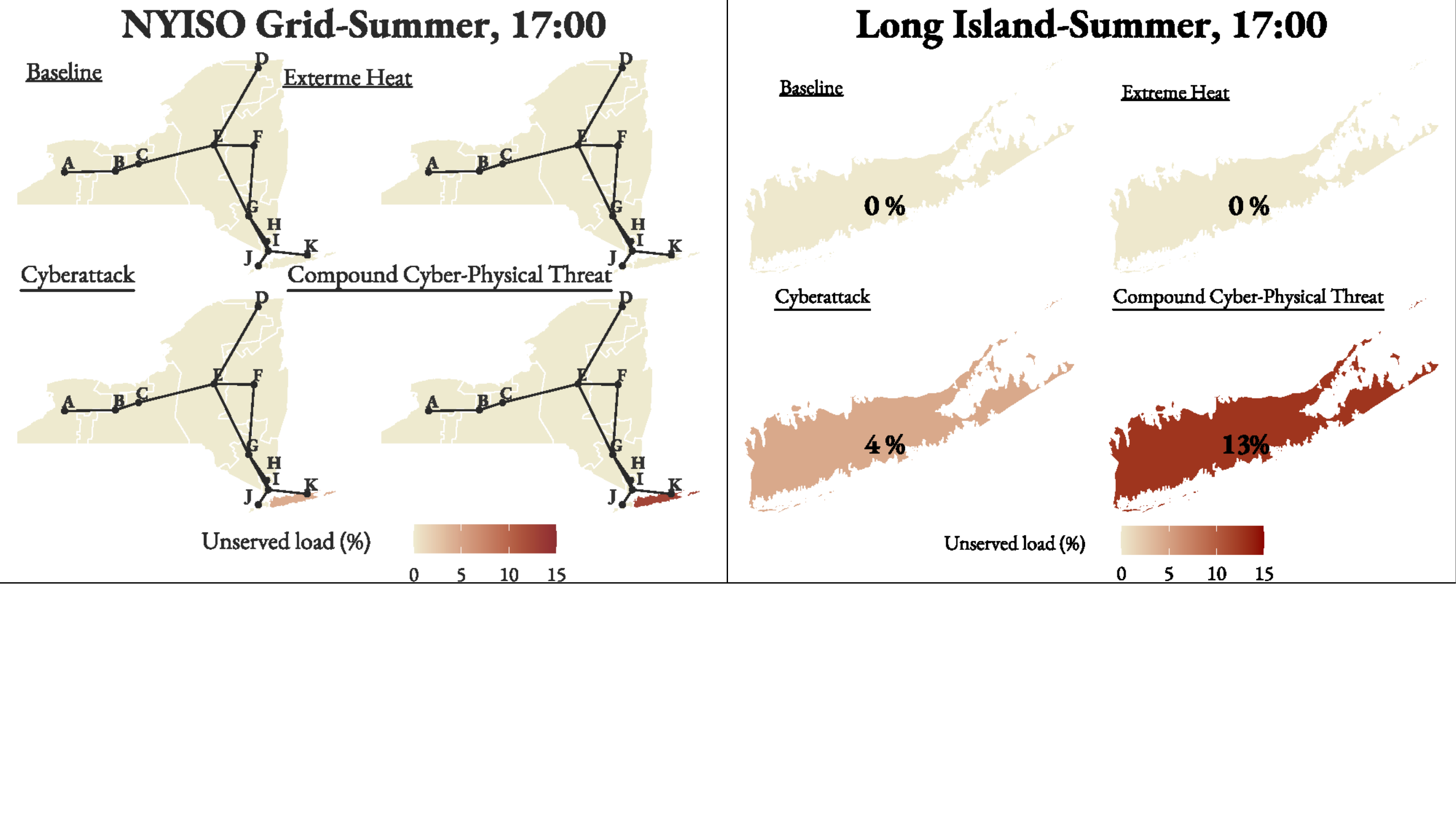}
		\caption{\label{fig:results-map}\footnotesize Regional NYISO Unserved Load (MW) at 17:00. Under the \textit{Heatwave} there is no unserved load. The \textit{Cyberattack} under normal operating conditions compromises 4\% of Long Island load and 1\% of NYISO load. The \textit{Compound Cyber-Physical Threat} increases load curtailment in Long Island to 13\%, while NYISO curtailment is  2\%.}
	\end{figure}

\textbf{Figure \ref{fig:results-map}} shows the regional distribution of unserved load in each scenario across all NYISO zones. We observe that the cyberattacker targets Long Island power generation in the \textit{Cyberattack} and \textit{Compound Cyber-Physical Threat} scenarios between 17:00-18:00PM, when the NYISO power network experiences the highest load shedding in both scenarios. The reason is that Long Island power generation is close to peak generation capacity at 18:00 on a summer day. Moreover, Long Island connects with only one other NYISO zone and the interconnection is close to capacity under the \textit{Baseline} scenario. Therefore, compromising power generation capacity has immediate impact on local unserved load.

    \begin{figure}[h!]
	    \centering	\includegraphics[width=1.0\textwidth]{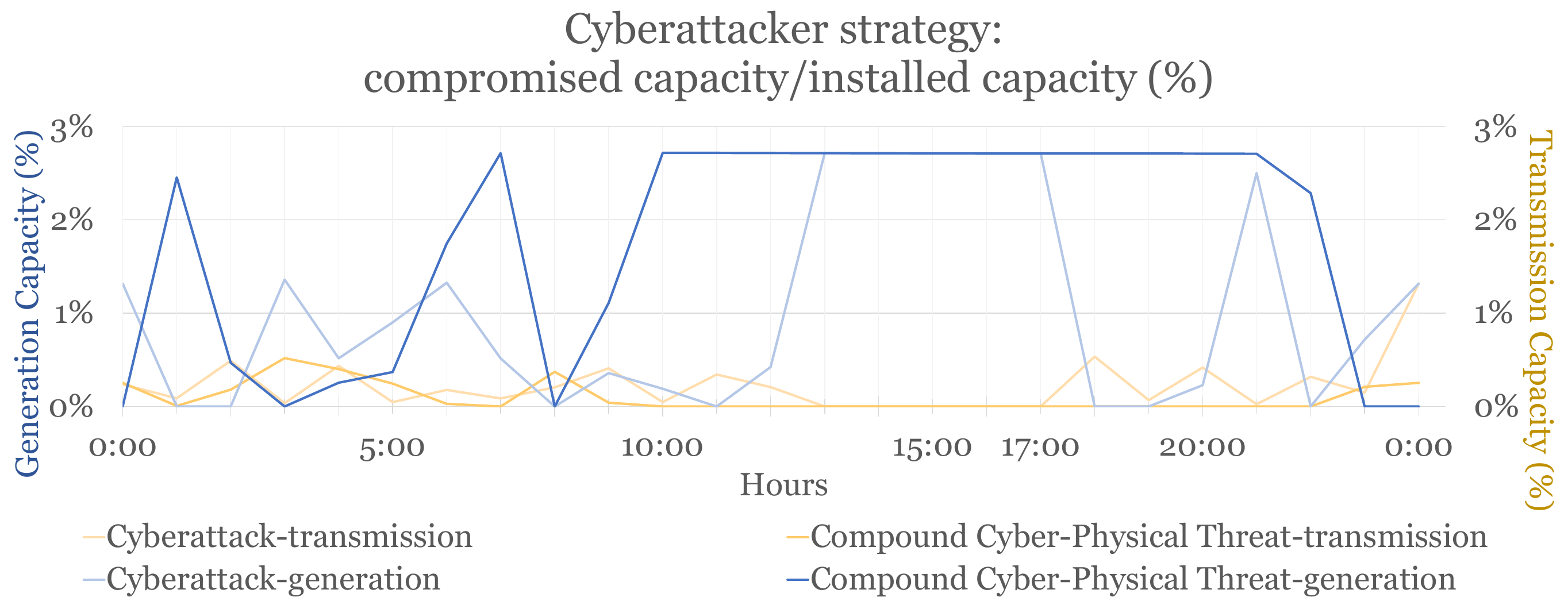}
		\caption{\label{fig:results-cyber-strategy}\footnotesize Cyberattacker strategy. Under the scenarios assumptions, the cyberattacker compromises consistently more generation capacity than transmission capacity. The cyberattacker allocates all of their budget to compromise 3\% of the total NYISO generation capacity during peak load shedding.}
	\end{figure}

Between compromising transmission and generation capacity in Long Island, which are operated close to their limits in the \textit{Baseline} scenario, the cyberattacker chooses to compromise more generation capacity.
\textbf{Figure \ref{fig:results-cyber-strategy}} shows the strategy of the cyberattacker. Between 13:00-18:00 in the \textit{Cyberattack} scenario, and 10:00-21:00 in the \textit{Compound Cyber-Physical Threat} scenario, when load curtailment happens in both scenarios, the attacker allocates all their resources to compromising generation capacity. \textbf{Figure \ref{fig:results-cyber-strategy}} reveals that the available cyberattacker resources allow them to compromise 3\% of total NYISO generation capacity during load curtailment periods. The optimal cyberattacker strategy arises from the distribution of cyberattacker costs across components, where we have assumed that compromising generation capacity is relatively cheaper compared to transmission capacity. However, our knowledge on the cyberattacker costs distribution and budget availability is limited and may change over time. For that, in the next section we test the sensitivity of our results to varying cyberattacker costs and budget availability.

\subsection{Infrastructure Vulnerabilities Sensitivity Analysis}\label{subsec:results-sensitivity}

\subsubsection{Cyberattacker cost distribution}\label{subsubsec:sensitivity-cybercost} In Section \ref{sec:scenario-design}, we describe how transmission components are often more protected against cyberattacks compared to consumer and generation components. Hence, in our formulation, we have assumed that the cost of compromising transmission capacity is disproportionately greater than the cost of compromising generation capacity for the same budget. In this Section, we explore the operational impact of a cyberattack if the difficulty of compromising transmission lines decreases in the future. For example, cybersecurity protocols may not keep up with the enhanced skillset of cyberattackers.
For that, a cyberattacker may be able to exploit novel attack vectors which can provide easier access to transmission components. In this section, we conduct sensitivity analysis on the relative cost $\left(\frac{\Tilde{c}^g}{\Tilde{c}^f}\right)$ compared to the default relative cyberattacker cost $\left(\frac{\Tilde{c}^{g,0}}{\Tilde{c}^{f,0}}\right)$, by decreasing the cyberattacker transmission cost by 10\% and increasing the cyberattacker generation cost by 10\% compared to their default values in each iteration, \textit{i.e.,} for iteration $i\in \mathcal{I}=\left\{1,2,\dots,6\right\}$, the relative cyberattacker cost $(\Tilde{c}^r)$ in each iteration is 

$$
    \Tilde{c}^r=\frac{\Tilde{c}^g}{\Tilde{c}^f} = \gamma^{(i)} \left(\frac{\Tilde{c}^{g,0}}{\Tilde{c}^{f,0}}\right), \qquad \gamma^{(i)}=\frac{1-(i-1)\cdot 0.1}{1+(i-1)\cdot 0.1},
$$

where $\gamma^{(i)}$ is the \textit{relative cyberattacker cost} ratio for iteration $i\in\mathcal{I}$.

\begin{figure}[h!]
	    \centering	\includegraphics[width=1.0\textwidth]{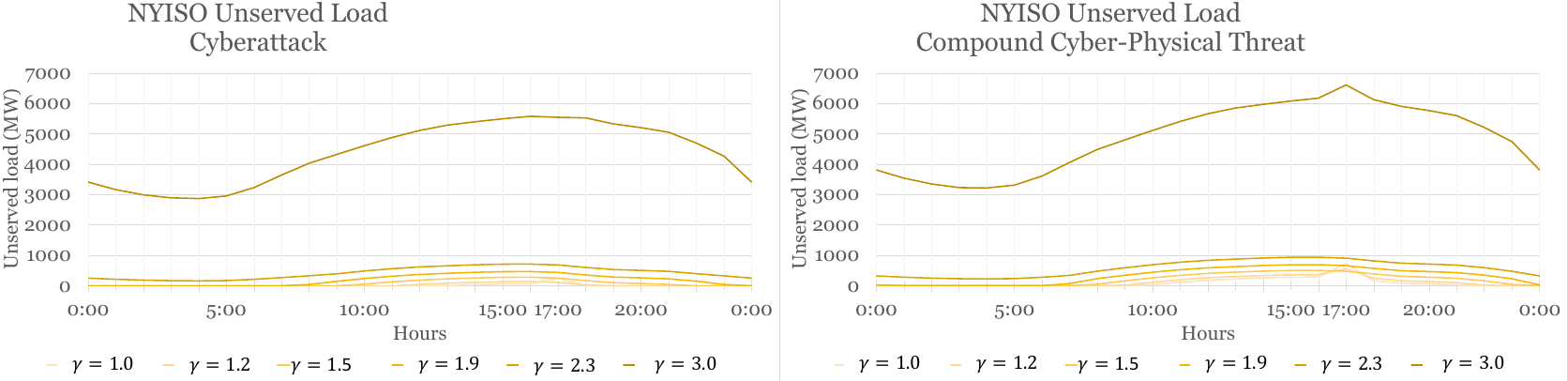}
		\caption{\label{fig:sensitivity-costs-unservedLoad}\footnotesize Unserved load during a summer 24-hour period for varying $\gamma$ under the \textit{Cyberattack} (left) and \textit{Compound Cyber-Physical Threat} (right) scenarios. Higher $\gamma$ increases the duration of unserved load and the impact for all hours and scenarios.}
	\end{figure}

\begin{figure}[h!]
	    \centering	\includegraphics[width=0.7\textwidth]{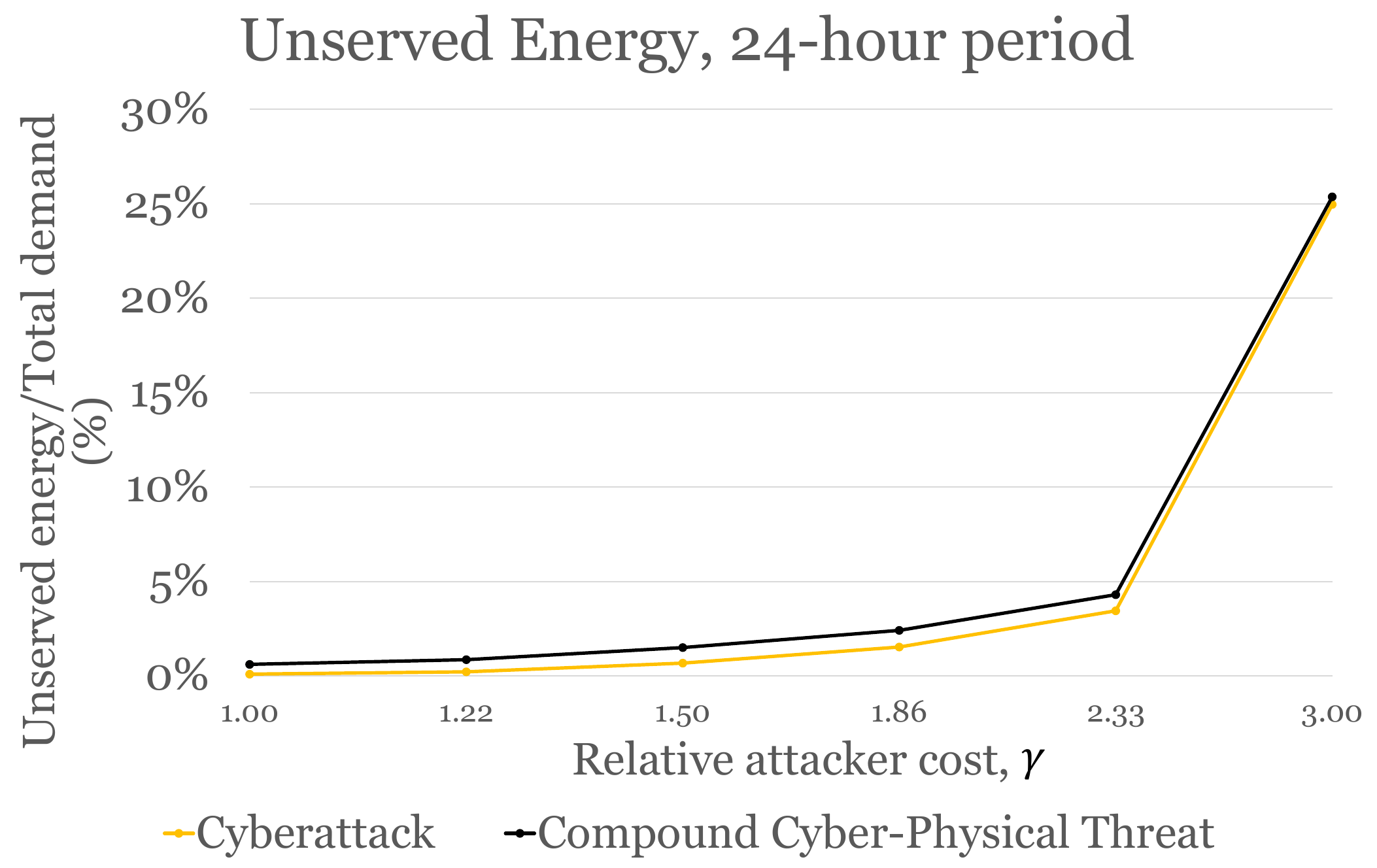}
		\caption{\label{fig:sensitivity-costs-unservedEnergy}\footnotesize Unserved electricity during a summer 24-hour period for varying relative cyberattacker cost ratio $(\gamma)$. Unserved electricity grows exponentially in the \textit{Compound Cyber-Physical Threat} and \textit{Cyberattack} scenarios. While unserved electricity is higher under the \textit{Compound Cyber-Physical Threat} compared to the \textit{Cyberattack}, the deviation between scenarios decreases for increasing $\gamma$.}
	\end{figure}

\begin{figure}[h!]
	    \centering	\includegraphics[width=1.0\textwidth]{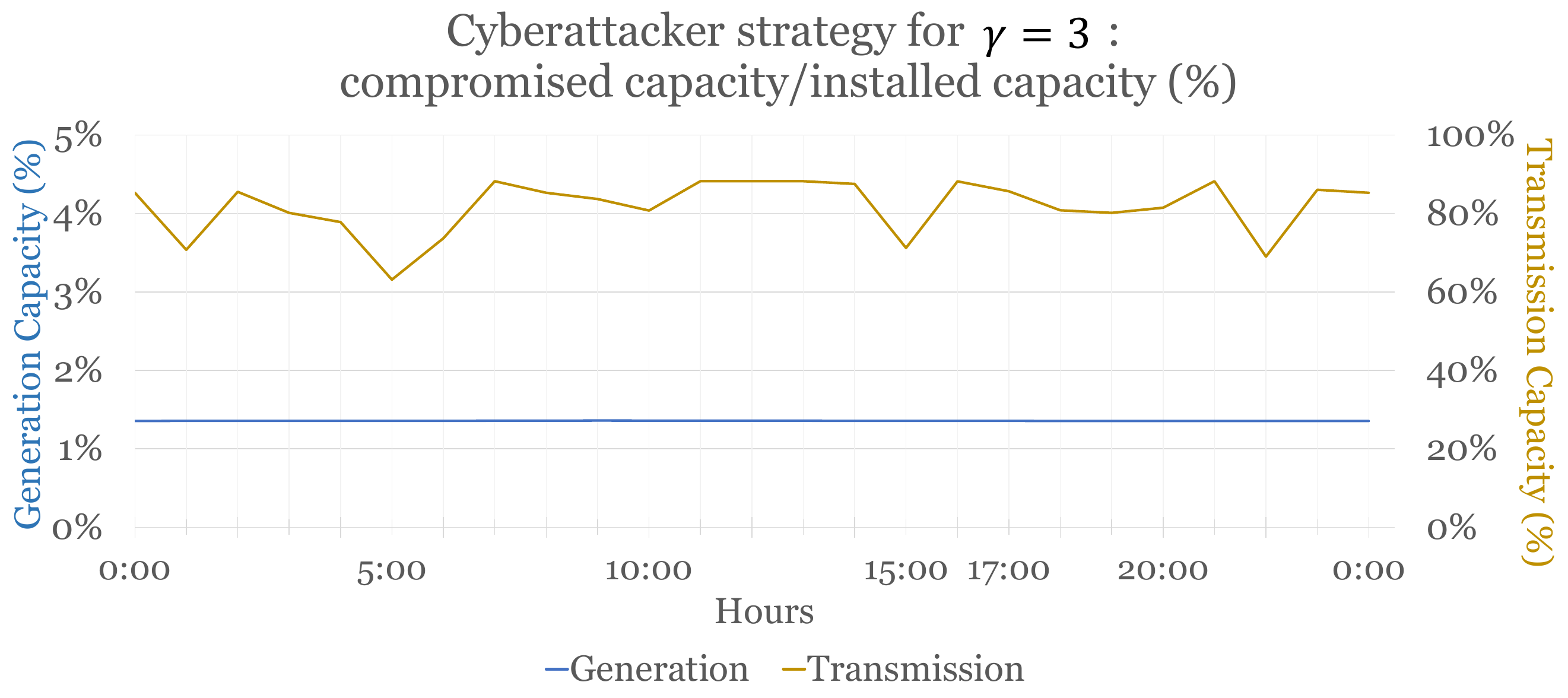}
		\caption{\label{fig:sensitivity-costs-strategy}\footnotesize Cyberattacker strategy for varying relative cyberattacker cost ratio $(\gamma)$. For $\gamma=3$, the cyberattacker cost of compromising transmission capacity is low enough to compromise more than 70\% of transmission capacity in all hours. The cyberattacker compromises 82\% of NYISO transmission capacity on average, while compromised generation capacity remains almost unchaged.}
	\end{figure}

\textbf{Figure \ref{fig:sensitivity-costs-unservedLoad}} shows that increasing $\gamma$ increases the impact and duration of load shedding. For $\gamma=1$, load shedding under the \textit{Compound Cyber-Physical Threat} happens  between 11:00-18:00 and peaks at 0.7 GW at 17:00, while for $\gamma=3$ load shedding spans all 24 hours and peaks at 6.6 GW. Unserved load in all hours is larger under the \textit{Compound Cyber-Physical Threat} than in the \textit{Cyberattack}. NYISO unserved electricity in a typical summer 24-hour period grows exponentially in $\gamma$ for the \textit{Cyberattack} and \textit{Compound Cyber-Physical Threat} scenarios, as shown in \textbf{Figure \ref{fig:sensitivity-costs-unservedEnergy}}. We also observe that greater $\gamma$ decreases the deviation of unserved electricity between the \textit{Cyberattack} and the \textit{Compound Cyber-Physical Threat} scenarios. A lower transmission attacker  cost allows for compromising more transmission capacity with the same budget. For a small enough cyberattacker transmission cost, the impact is extremely severe, up to 25\% of unserved electricity for $\gamma=3$, or 4.95 million customers, under both scenarios compared to no more than 1\% of unserved electricity for $\gamma=1$, or 198,000 customers. 
Therefore, for $\gamma \geq 3$, the impact of the \textit{Compound Cyber-Physical Threat} depends vastly on the capabilities of the capabilities. Finally, the cyberattacker strategy in \textbf{Figure \ref{fig:sensitivity-costs-strategy}} provides more insights on the exponential growth  of unserved electricity for increasing $\gamma$. When $\gamma=3$, compromising transmission capacity is cheap enough for the cyberattacker to compromise more than 70\% of NYISO transmission capacity and an average of 82\% of NYISO transmission capacity each hour under both the \textit{Cyberattack} and  \textit{Compound Cyber-Physical Threat} scenarios. By attacking transmission capacity, the Genesee zone disconnects from neighboring zones. When Genesee is disconnected from neighboring zones, compromised Genesee generation capacity has direct impact on unserved load in Genesee. Therefore, the cyberattacker chooses to also compromise 500 MW of Genesee generation capacity in all hours, or 1\% of NYISO installed capacity, leading to 46\% of unserved load in Genesee.

\begin{figure}[h!]
	    \centering	\includegraphics[width=0.7\textwidth]{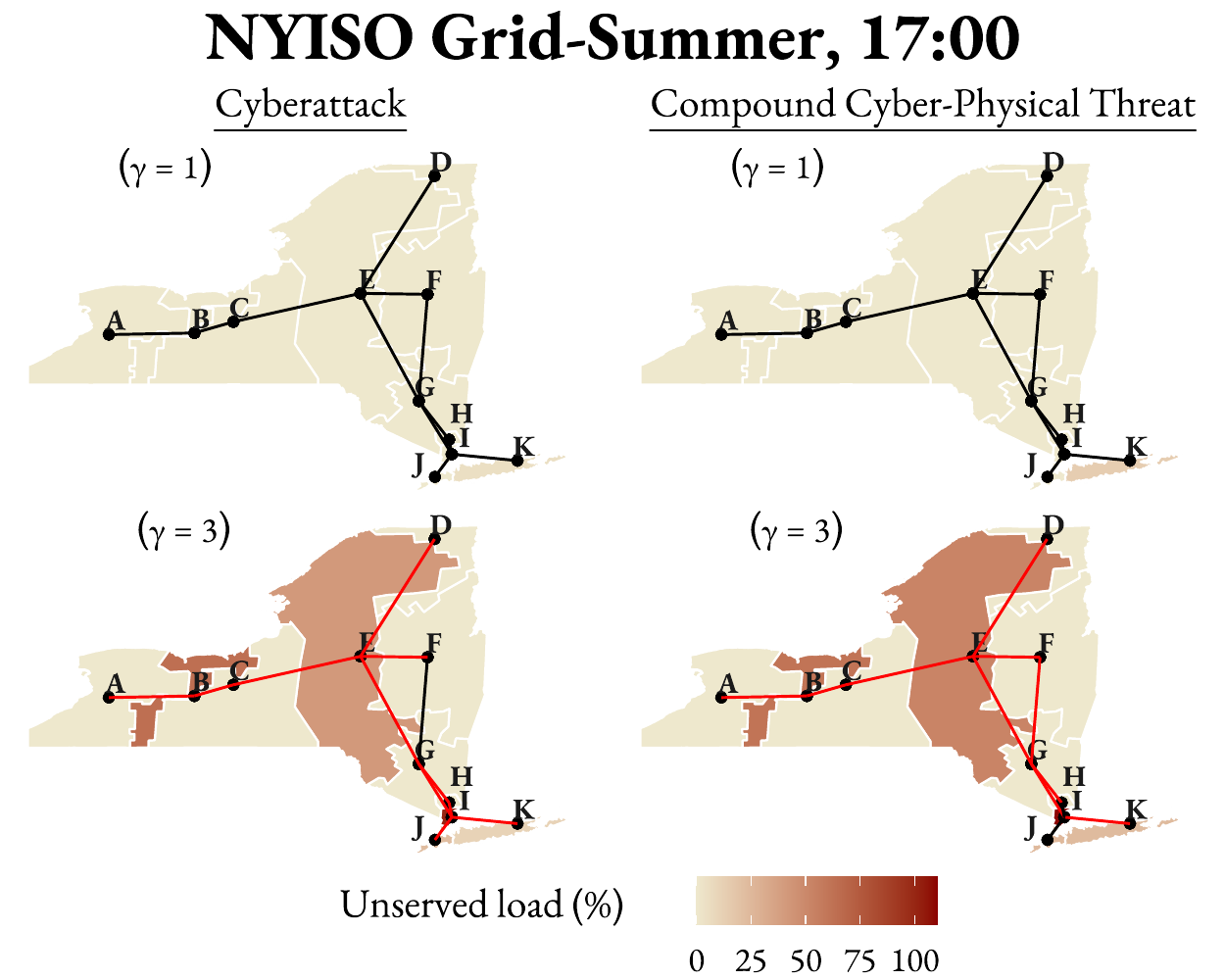}
		\caption{\label{fig:sensitivity-costs-map}\footnotesize Regional distribution of unserved load within the NYISO grid for a summer day, at 17:00. The cyberattacker compromises less than 5\% and more than 95\% of transmission capacity for black and red lines respectively. tha For $\gamma=3$, the attacker compromises two additional interconnections compared to the \textit{Cyberattack} scenario between Capital (F) and Hudson Valley (G), and Dunwoodie (I) and New York City (J), leading to a 10\% increase in New York City unserved load during peak load shedding. }
	\end{figure}

\textbf{Figure \ref{fig:sensitivity-costs-map}} shows the spatial distribution of load curtailment within NYISO during a typical summer day for the time-period with longest load curtailment. Electricity importing regions, \textit{i.e.,} regions whose average electricity demand exceeds regional installed capacity, are the most affected. For example, the Genesee (B), Mohawk (E), and Dunwoodie (I) regions face 56\%, 48\%, and 100\% load curtailment during the peak curtailment period at 17:00. Finally, for $\gamma=3$, the \textit{Cyberattack} timed with the \textit{Heatwave} increases unserved load in New York City by 10\% compared to the \textit{Cyberattack} under normal operating conditions. In the \textit{Compound Cyber-Physical Threat} scenario, the attacker compromises transmission capacity between Capital (F) and Hudson Valley (G), and Dunwoodie (I) and New York City (J), which isolates New York City from upstate supply from Capital (F) and Hudson Valley (G). Unserved load increases from 2 GW in the \textit{Cyberattack} scenario to 2.2 GW in the \textit{Compound Cyber-Physical Threat} scenario.

\subsubsection{ Availability of Resources to the Cyberattacker}\label{subsubsec:sensitivity-budget}
In this section we explore the sensitivity of our results to the availability of cyberattacker resources. We use the default cyberattacker costs defined in Section~\ref{sec:scenario-design}, but increase the budget $(\Tilde{b})$ by 20\% compared to the default budget $(\Tilde{b}^0)$ in each iteration $i\in\mathcal{I}$, \textit{i.e.,}
$$
    \Tilde{b} = \beta^{(i)} \cdot b^0, \qquad \beta^{(i)}=(1+(i-1)\cdot 0.2),
$$

where $\beta^{(i)}$ is the \textit{relative cyberattacker resources} ratio for iteration $i\in\mathcal{I}$.

\begin{figure}[h!]
	    \centering	\includegraphics[width=1.0\textwidth]{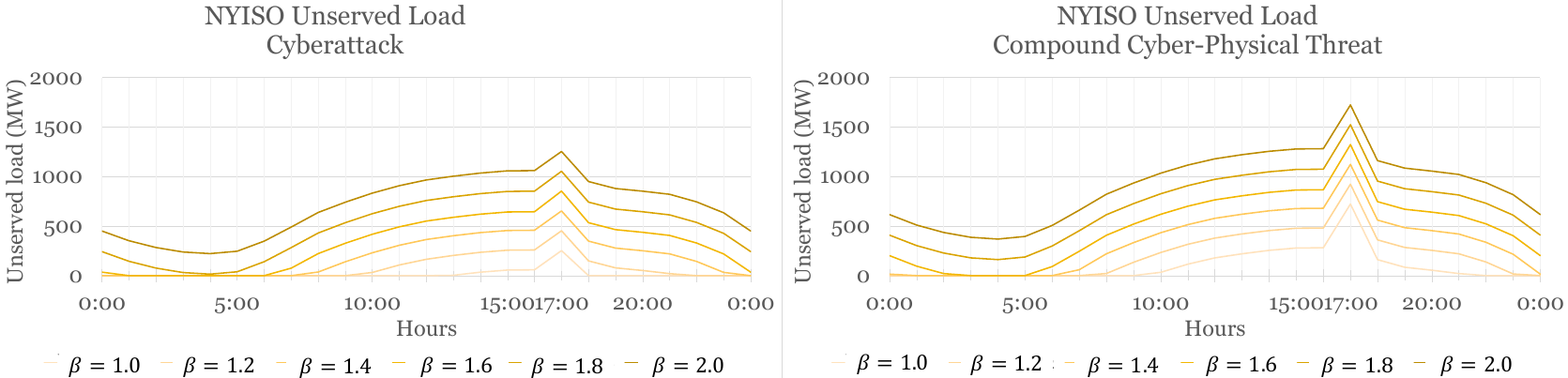}
		\caption{\label{fig:sensitivity-budget-unservedLoad}\footnotesize Unserved load during a summer 24-hour period for varying $\beta$ under the \textit{Cyberattack} (left) and \textit{Compound Cyber-Physical Threat} (right) scenarios. Higher $\beta$ increases the duration of unserved load and the impact for all hours and scenarios.}
	\end{figure}

\begin{figure}[h!]
	    \centering	\includegraphics[width=0.7\textwidth]{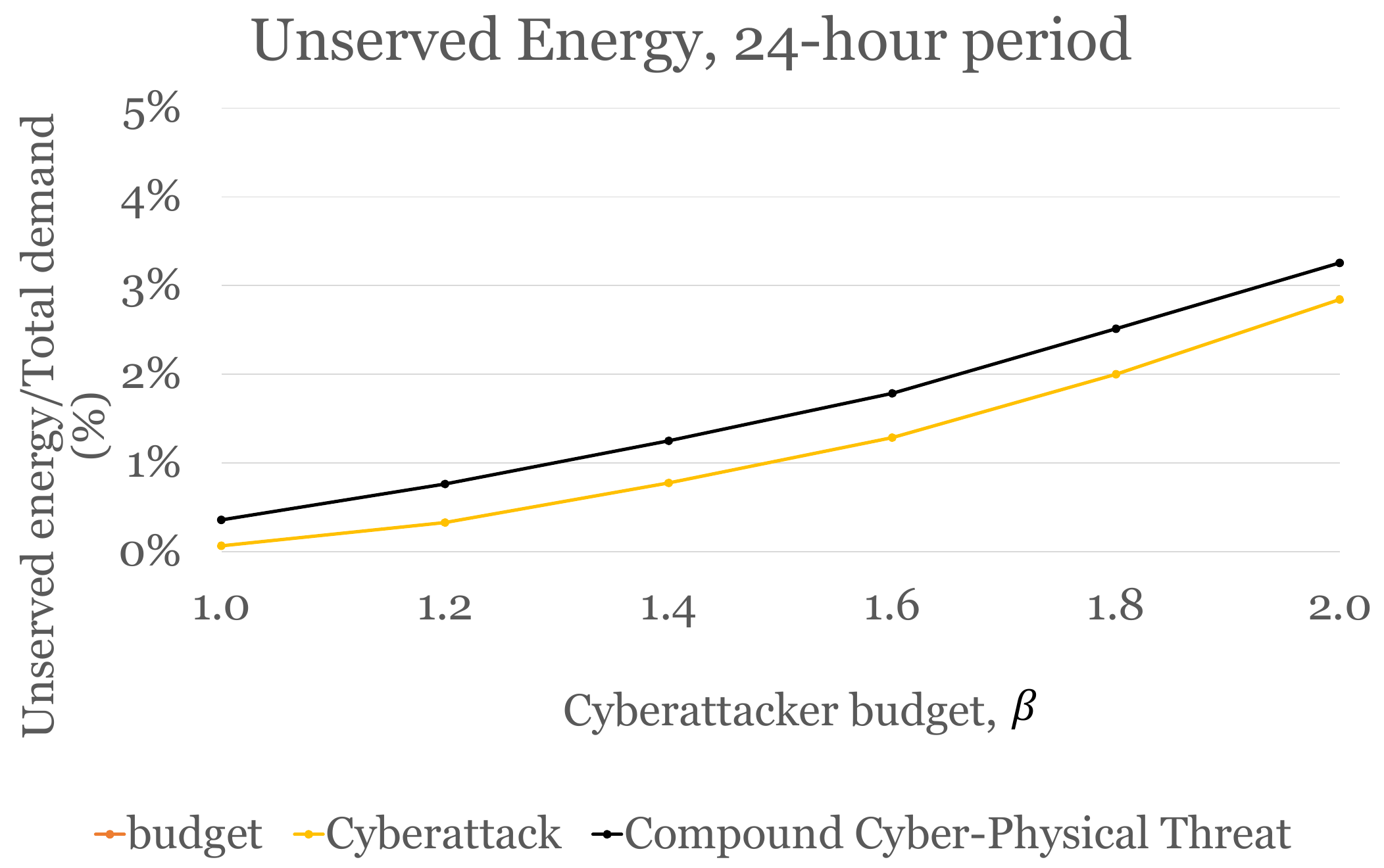}
		\caption{\label{fig:sensitivity-budget-unservedEnergy}\footnotesize Unserved electricity during a summer 24-hour period for varying relative cyberattacker resources ratio $(\beta)$. Unserved electricity grows linearly in the \textit{Compound Cyber-Physical Threat} and \textit{Cyberattack} scenarios. Unserved electricity is higher under the \textit{Compound Cyber-Physical Threat} compared to the \textit{Cyberattack} and grows by 3\% in both scenarios when the cyberattacker resources double with respect to the default budget ($\beta=2$).}
	\end{figure}

\begin{figure}[h!]
	    \centering	\includegraphics[width=1.0\textwidth]{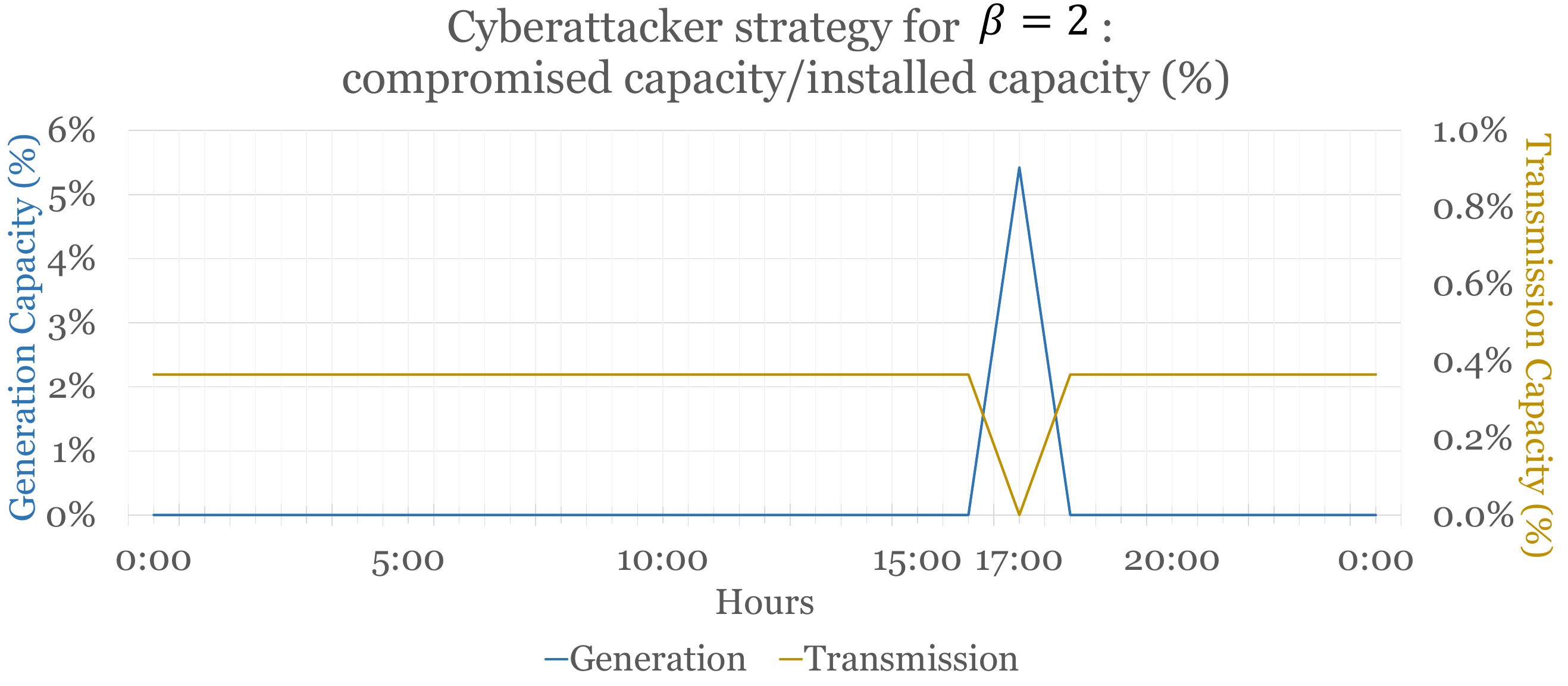}
		\caption{\label{fig:sensitivity-budget-strategy}\footnotesize Cyberattacker strategy for varying relative cyberattacker resources ratio $(\beta)$. For double resources ($\beta=2$), the cyberattacker chooses to allocate their budget between compromising transmission and generation capacity in each hour similarly to the default budget scenario ($\beta=1$), \textit{i.e.,} compromise generation capacity during the peak load shedding period.}
	\end{figure}

\textbf{Figure \ref{fig:sensitivity-budget-unservedLoad}} shows that increasing the cyberattacker budget $\Tilde{b}$ increases the impact and duration of load shedding. For $\beta=1$, load shedding under the \textit{Compound Cyber-Physical Threat} happens  between 11:00-18:00 and peaks at 0.7 GW at 17:00, while for $\beta=2$ load shedding spans all 24 hours and peaks at 1.7 GW. Unserved load in all hours is greater under the \textit{Compound Cyber-Physical Threat} compared to the Cyberattack. NYISO unserved electricity in a typical summer 24-hour period grows linearly in $\beta$ for the \textit{Cyberattack} and \textit{Compound Cyber-Physical Threat} scenarios, as shown in \textbf{Figure \ref{fig:sensitivity-budget-unservedEnergy}}. Greater $\beta$ preserved the percentage deviation of unserved electricity between the \textit{Cyberattack} and the \textit{Compound Cyber-Physical Threat} scenarios. \textbf{Figure \ref{fig:sensitivity-budget-strategy}} shows that the cyberattacker chooses a similar strategy as in Section \ref{subsec:results-default}, but can cause greater damage due to the availability of additional resources. The cyberattacker can cause similar damage by targeting transmission and generation capacity for all time periods except for 17:00, hence the choice to compromise transmission capacity. During the peak electricity demand hour, close to 17:00, the cyberattacker can cause additional unserved load by targeting capacity. Hence, during that hour, the cyberattacker focuses all their resources on generation capacity. When the cyberattacker doubles their budget, unserved load increases by 3\%, or roughly 594,000 customers, in both scenarios.

\begin{figure}[h!]
	    \centering	\includegraphics[width=1.0\textwidth]{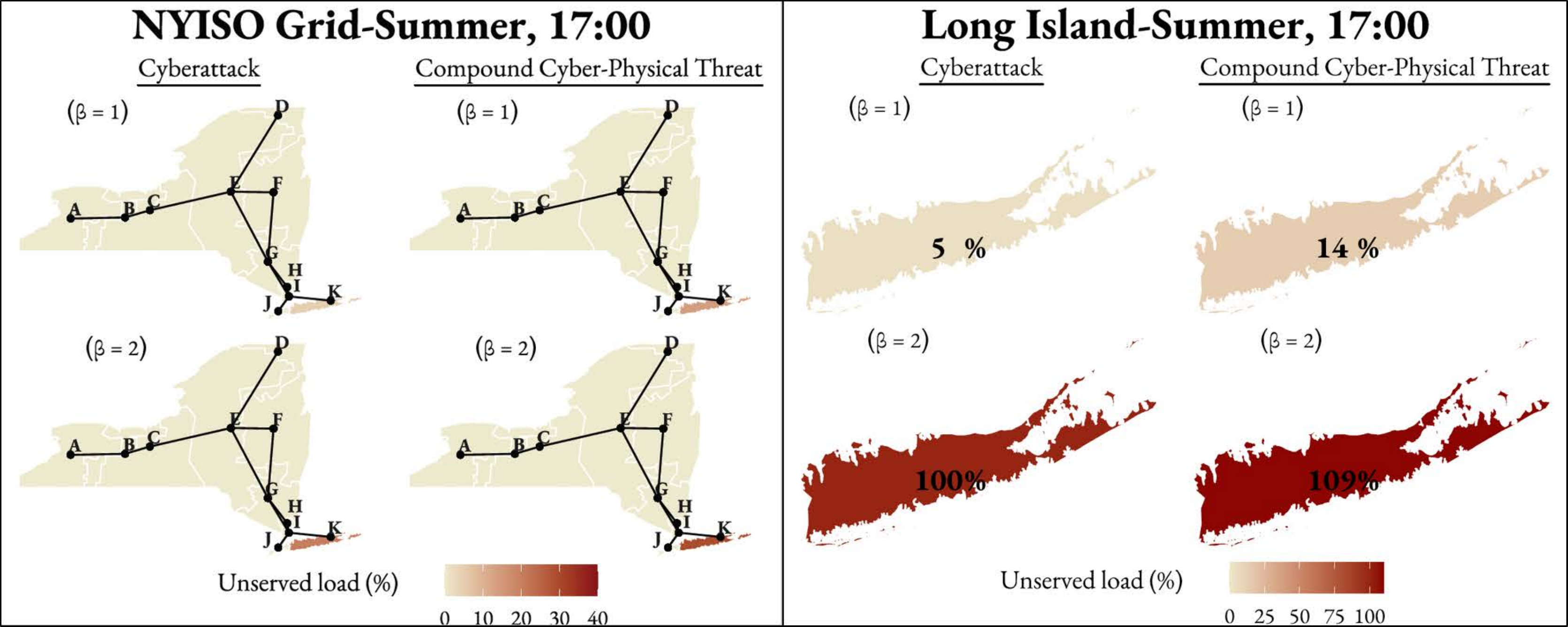}
		\caption{\label{fig:sensitivity-budget-map}\footnotesize Regional distribution of unserved load within the NYISO grid for a summer day, at 17:00. For $\beta=2$, the attacker has the capabilities to compromise enough generation and transmission capacity in Long  Island to and cause a blackout in the whole region during normal operating conditions and during the \textit{Heatwave} scenarios.}
	\end{figure}

\textbf{Figure \ref{fig:sensitivity-budget-map}} reveals that the spatial distribution of unserved load remains insensitive to increasing the budget, but the magnitude of unserved demand is magnified. Although the total curtailed electricity in the 24-hour period increases to 3\% when doubling the budget, Long Island unserved load increases to 100\% under the \textit{Cyberattack} and 109\% under the \textit{Compound Cyber-Physical Threat} compared to the \textit{Baseline}. The \textit{Cyberattack} scenario results show that  doubling the budget of the cyberattacker can compromise all of Long Island generation. In the \textit{Compound Cyber-Physical Threat}, the system operator sheds 109\% of the \textit{Baseline} load, \textit{i.e.,} the load under normal operating conditions, and the additional 9\% of the \textit{Heatwave}-induced demand increase.

\subsection{Economy-Wide Impacts}

\begin{figure}[h!]
	    \centering	\includegraphics[width=1.0\textwidth]{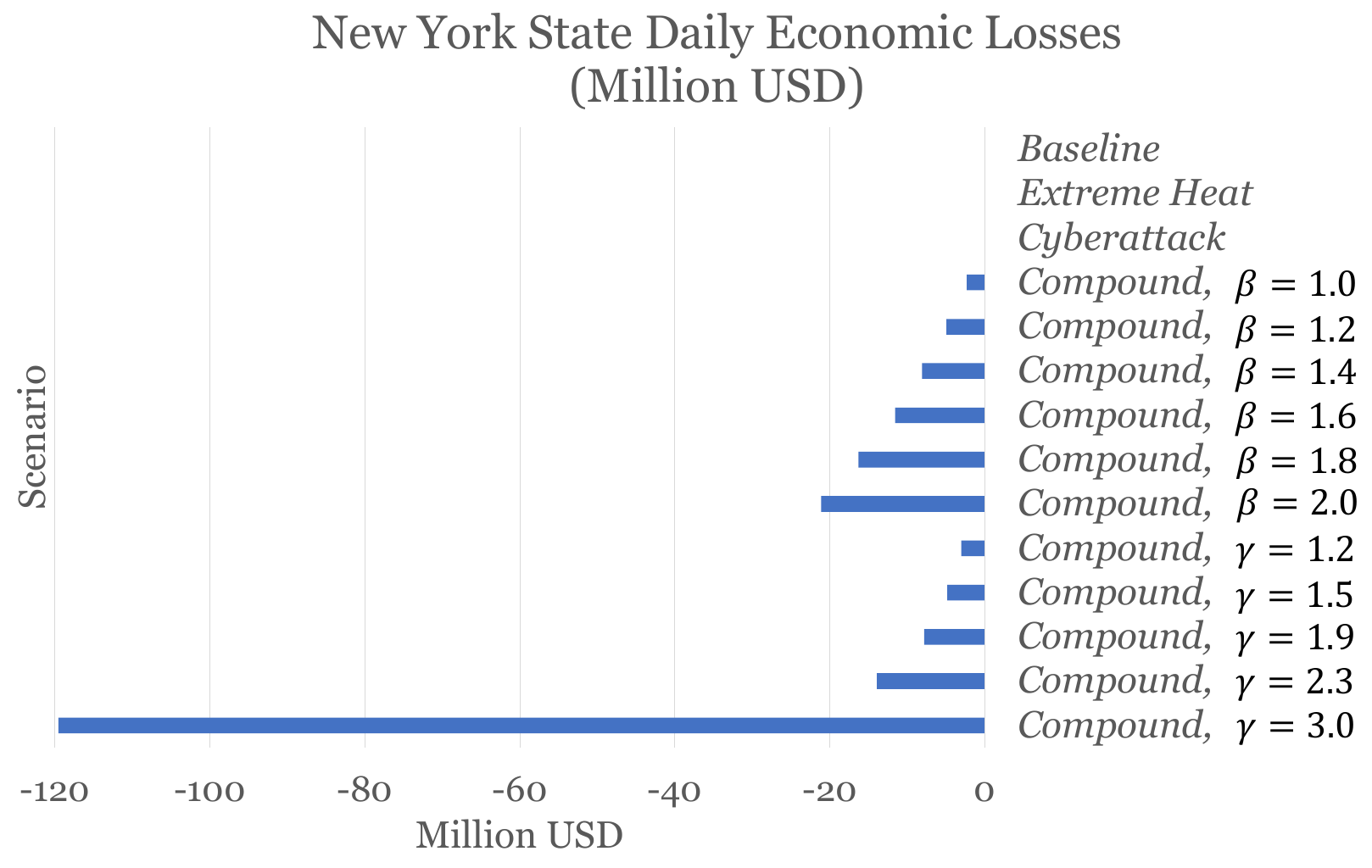}
		\caption{\label{fig:macro-losses}\footnotesize Economic losses measured as decrease of daily GDP compared to \textit{Baseline} GDP. Doubling the cyberattacker resources aggravates economy-wide losses by 10 times, while enhanced transmission vulnerability can lead to a 60-fold increase of economy-wide losses.}
	\end{figure}

In this section, we evaluate the ripple effects of the \textit{Compound Cyber-Physical Threat} across all economic sectors using the WiNDC model. 
In Figure \ref{fig:macro-losses} we can observe the compounding nature of a timed \textit{Cyberattack} and a \textit{Heatwave}. Under the \textit{Cyberattack} and \textit{Heatwave} scenarios there is no or little unserved load respectively, similarly to the \textit{Baseline} scenario, and economic losses are negligible. However, under the \textit{Compound Cyber-Physical Threat}, the New York State economy incurs \$2 Million losses. Increasing the cyberattacker budget magnifies economic losses, up to \$21 Million for a budget twice the default cyberattacker budget. Under the \textit{Compound Cyber-Physical Threat}, doubling the cyberattacker budget causes a 9-fold increase of unserved load in the electric power sector, from 2.4 GW to 21.8 GW, and a 10-fold increase of economy-wide losses. Similarly, increasing the vulnerability of transmission components to cyberattacks $(\gamma)$ increases economy-wide losses exponentially to \$120 Million for $\gamma=3$, \textit{i.e.,} 2.3\% of the daily GDP. 

\begin{figure}[h!]
	    \centering	\includegraphics[scale=0.7]{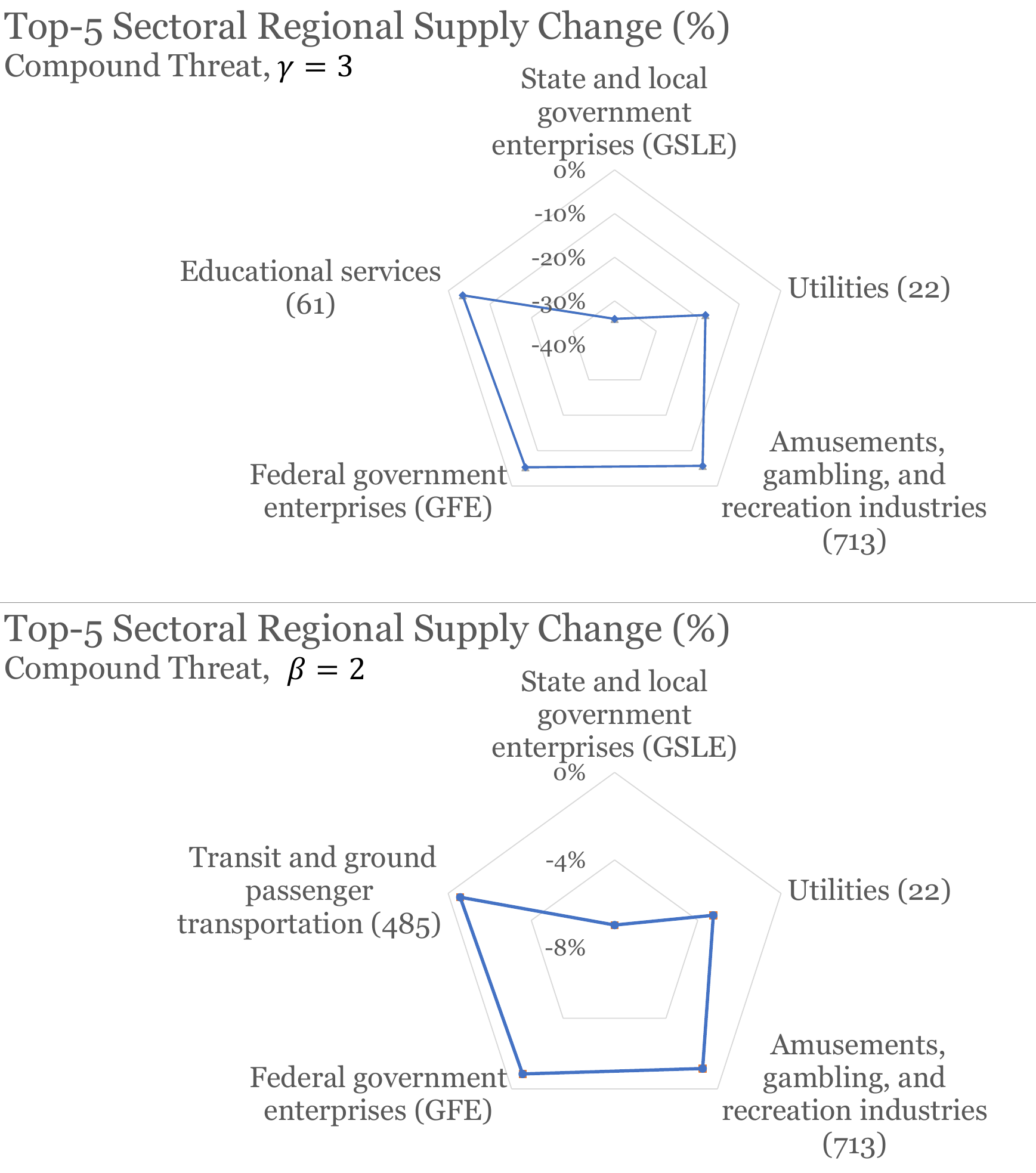}
		\caption{\label{fig:macro-top5}\footnotesize Top-5 most impacted sectors under the \textit{Compound Cyberphysical Threat} scenario with $\gamma=3$ (up) and $\beta=2$ (down). The most affected sector is the \textit{``State and local government enterprises (GSLE)"} in both cases.}
	\end{figure}

The economic impact can vary across sectors. From a total of 71 sectors, only 5 sectors report losses greater than 0.5\% under the \textit{Compound Cyber-Physical Threat} with $\beta=2$. When $\gamma=3$, 16 sectors report losses greater than 0.5\%. Figure \ref{fig:macro-top5} shows that the impact varies even within the top-5 most impacted sectors under the \textit{Compound Cyber-Physical Threat}. The \textit{``State and local government enterprises"} face 7\% losses under the \textit{Compound Cyber-Physical Threat} with $\beta=2$ and 34\% losses when $\gamma=3$. The top-4 sectors are the same for both scenarios and include also the \textit{``Federal government enterprises (GFE)"} and \textit{``Amusements, gambling, and recreation industries"}, with average losses of 1\% when $\beta=2$ and 5\% when $\gamma=3$. The \textit{``Educational services"} sector completes the top-5 with 3\% for $\gamma=3$ and the ranks 5th with 1\% for $\beta=2$.

\section{Discussion}\label{sec:discussion}
Our results reveal which electric power infrastructure components are most vulnerable to cyberattacks under extreme weather stressors. Depending on the budget of the attacker, unserved load in the NYISO increases from 0 to 3\% as the budget doubles, however the impact is highly localized. Less interconnected NYISO zones are heavily targeted, leading to disproportionally more unserved load compared to more interconnected zones with greater local electricity supply, which remains relatively unaffected.

\subsection{Electric Power Network Resilience and Economy-Wide Response}

\paragraph{\textit{The cyberattacker targets nodes of the NYISO electric power network with low network connectivity}.} The Long Island and New York City zones have the second and third greatest installed capacity. Moreover, the aggregate Long Island and New York City demand absorbs all of Long Island and New York City electricity production respectively. The two zones are similar in that they connect with other NYISO zones via relatively low-capacity interconnections. Therefore, compromising generation and transmission capacity in these two nodes translates directly to unserved load. Note that excess transmission capacity in the Long Island zone implies that the cyberattacker needs to compromise a larger portion of NYISO transmission capacity, which is costlier compared to compromising generation capacity, to cause unserved load. Thus, between the two zones with low connectivity, the cyberattacker compromises the Long Island zone first.

\paragraph{\textit{For each targeted node, the cyberattacker compromises generation capacity before compromising power flow capacity}.} In Section \ref{sec:results}, we observed that the first targeted zone was Long Island. For example, under default cyberattacker costs and budget, \textit{i.e.,} $\beta=1$ and $\gamma=1$, at 17:00, the cyberattacker compromises 20\% of Long Island generation capacity. As the attacker budget doubles, the cyberattacker compromises more generation capacity. In the case of Long Island and New York City, the cyberattacker prefers compromising generation capacity before transmission capacity for two reasons. First, compromising generation capacity requires less resources. In Section \ref{sec:scenario-design} we assume that power generation follow less rigorous cybersecurity protocols, compared to the transmission system, hence are more vulnerable to cyberattacks. For that, a cyberattacker needs less resources to compromise a MW of generation capacity compared to transmission capacity. Second, the low connectivity of both zones with the other NYISO zones implies that the two zones have less access to excess capacity within the NYISO grid. Given their low connectivity and that the Long Island and New York City zones cover almost all of their load from local production, compromising generation capacity has a direct impact on unserved load and the cyberattacker targets the Long Island zone first.

\paragraph{\textit{Increasing the vulnerability of transmission components increases unserved load exponentially, while increasing the cyberattacker resources availability increases unserved load linearly.}} We model increasing  vulnerability of transmission components by decreasing the cyberattacker transmission cost, which allows the cyberattacker to compromise more transmission capacity for the same budget. For $\gamma=3$, the cost of compromising transmission capacity is so low, that the cyberattacker can compromise most of the NYISO transmission capacity. On average, the cyberattacker compromises 82\% of transmission capacity in the 24-hour period for $\gamma=3$. When $\gamma=3$, the cyberattacker can compromise more than 70\% of transmission capacity, which leaves most zones largely dependent on their domestic generation. When transmission components vulnerability is lower, the attacker can compromise less transmission capacity, due to their resource constraint, and excess transmission capacity allows the NYISO to absorb the impact. However, under high vulnerability, significant transmission capacity is compromised, and additional losses in transmission capacity lead to unserved load in multiple zones. For example, compromising transmission between Hudson Valley (G) and Dunwoodie (I) disconnects most major upstate generators from supplying New York City and Long Island demand. On the other hand, increasing the budget under default cyberattacker costs leads to targeting generation components. However, excess generation capacity in the NYISO grid and the ability to substitute zonal generation with imports from other NYISO zones leads to an increase of 19.3 GWh in unserved load, affecting more than 594,000 customers, for double the cyberattacker resources.

\paragraph{\textit{\textit{A Compound Cyber-Physical Threat compounds also economy-wide losses}}.} The lack of regional electricity supply constrains the activity of other sectors. Across all economic sectors, the \textit{Heatwave} and \textit{Cyberattack} do not impact other sectors, while a \textit{Compound Cyber-Physical Threat} can decrease economic activity by \$2 million. We observe the largest impact for the largest operational disruption when $\gamma=3$, which leads to losses of \$120 million or 2.7\% of daily New York State GDP.  Beyond the ``Utilities" sector, we identify the ``Federal government enterprises", ``State and local enterprises" and ``Transit and ground passenger transportation" sectors as particularly vulnerable to disruptions in electricity supply in New York State. These enterprises provide essential services for local well-being,  \textit{e.g.,} transportation. Therefore, disruptions in the activity of these three sectors can impact a broad range of population groups. The assumptions underpinning a CGE, discussed in the next Section, constrain the quantification of economy-wide losses. Our analysis does not aim to quantify economy-wide losses or the VOLL, which requires extensive \textit{ex-post} surveys. Our goal is to provide an \textit{ex-ante} computation of the lower bound, meaning that the impact of the \textit{Compound Cyber-Physical Threat} to New York State GDP and individual sectors will be \textit{at least} as high as in our analysis. For example, across scenarios, we derive losses between \$1,054-\$1,078 per MWh in the 24-hour period. For comparison, the estimated losses for other than compound cyber-physical stressors, e.g.,  26-hour long 1977 New York City blackout and  48-hour long 2003 Northeastern blackout, were \$3,405 and \$9,300 per MWh, respectively \citep{abidi2015}. The response of the economy is the result of the microeconomic principles underpinning CGEs, hence the results serve as an indicator of sectors which are more likely to face disruptions.

\subsection{Limitations} 
The limitations of our study are inherent to the nature of the \textit{Compound Cyber-Physical Threat} against critical infrastructures and the computational capabilities of existing methodologies. 

First, actual data on features of the transmission and distribution system are confidential \cite{birchfield2016}. Hence, our results are sensitive to electric power network features, which in our case include transmission capacities, voltage angle difference limits, and generation capacities.


Second, our results are sensitive to the assumptions on the cyberattacker's rationale. In this paper, we assume scenarios that are rapid and have immediate, localized impact on consumers. Alternatively, a cyberattacker may opt for longer-term strategies to exploit less severe events. For example, a cyberattacker may want to maximize the operating cost of the NYISO. In this case, the cyberattacker may attempt to inflict less severe damage such that they go unnoticed. Moreover, under a scenario of NYISO cost maximization, the cyberattacker may opt to compromise low-cost resources first in order to force the NYISO to dispatch more expensive fuel-fired generators into the fuel mix more often. When factoring in interdependent infrastructures, the cyberattacker may decide to compromise power supply to zones with critical telecommunications components or wastewater treatment facilities. Our framework allows the flexibility to adjust the cyberattacker objectives $\tilde{C}(\cdot)$. Moreover, the MCP reformulation of the lower-level problem can be enhanced to account for interdependencies with other critical sectors, \textit{e.g.,} the water/ wastewater sector.


Third, our results are sensitive to the elasticities of substitution between production inputs of CGE sectors. In Section~\ref{subsec:windc} we mentioned that CGEs have yearly time steps and thus  input substitutions happen at a longer time-scale compared to the 24-hour time-frame studied here. However, it may be more difficult to substitute between inputs in the short-term, \textit{e.g.,} between different energy sources. In addition, post-disruption elasticities of substitution are not available, even for individual sectors. Hence, our analysis is not suitable for the quantification of economic losses. Instead, we aim to identify vulnerable sectors and their response, and provide an \textit{ex-post} calculation of a lower bound on economic losses which is consistent in the orders of magnitude.


\section{Conclusions}\label{sec:conclusions}

\subsection{Summary and Contributions}
This paper introduces the novel \textit{Compound Cyber-Physical Threat}, arising when a cyberattack targets electric power network components when they are critically-stressed due to extreme weather events. Our work contributes to the identification and management of electric power network vulnerabilities, and economy-wide losses. We find that when generators are more easily compromised, the cyberattcker targets NYISO zones with low network connectivity. While a \textit{Cyberattack} under normal operating conditions or a \textit{Heatwave} barely disrupt the NYISO electric power network, a \textit{Compound Cyber-Physical Threat} exacerbates regional unserved electricity demand under the individual threats by more than three times, leading to 13\% unserved load in a 24-hour summer period. When transmission lines are easier to compromise, the cyberattacker targets NYISO zones with low network connectivity and a lack of generation capacity. The \textit{Compound Cyber-Physical Threat} exacerbates also economy-wide losses across sectors. The individual \textit{Heatwave} and \textit{Cyberattack} scenarios result in an insignificant GDP reduction. However, depending on the cyberattacker cost of each strategy, GDP reduction ranges between 0-2.7\% of daily GDP.

In the context of cyberattacks timed with extreme weather events, we introduce a framework that accounts for the operational capabilities of a cyberattacker and test the sensitivity of our framework against the cyberattacker resource availability. In addition, our framework couples the cyberattacker strategy with the response of all sectors in the economy. We also contribute to the derivation of a lower bound of economy-wide losses at the state level. We show that our estimation is consistent in orders of magnitude with estimates of losses in past blackouts. Although the quantification of economy-wide losses is beyond the capabilities of our framework, further refinement of the proposed methodology can contribute to estimating a \textit{``reference point"} for the VOLL under \textit{Compound Cyber-Physical Threats}. Therefore, further development of our framework and realistic data can assist system operators and federal, state and local agencies in infrastructure planning against extreme weather events and cyberattacks. 

\subsection{Future Research}

This paper focuses on heatwaves, but our framework is applicable to emerging, billion-dollar extreme weather events, including hurricanes, storms, and wildfires. Hurricanes, storms, and wildfires follow a trajectory, thus disproportionately affecting infrastructures across the trajectory compared to the rest of the system. Therefore, their impact is highly localized. Understanding the regional impact of extreme weather events, also across communities with varying sociodemographic characteristics, requires a methodology with greater spatial resolution. Hence, future research should model both the distribution and the transmission system in a single model.
Moreover, the analysis of distribution-level impacts requires a county-level CGE
dedicated to extreme events. Such CGE requires recalibrating the sectoral inputs substitution elasticities to account for the short-term effects during and following the extreme event. 

Second, this paper focuses on electric power networks, due to their critical contribution to the operation of most sectors. However, a disruption in the electricity sector can propagate to multiple critical infrastructures through their interdependencies and further compound the impact of an extreme event on regional infrastructures and economic activity. Therefore, going beyond operational disruptions and accounting for the public health impacts of \textit{Compound Cyber-Physical Threats} requires understanding the propagation of failures in the water/wastewater, telecommunications, and transportation infrastructures.







\section{Acknowledgements}

The authors are grateful to Dr. Maxwell Brown and Dr. Jon Becker for their generous advice on the implementation of WiNDC. 

\bibliographystyle{apalike}
\bibliography{main}

\newpage
\section*{Supplementary Material: Abbreviations and Nomenclature}\label{apdx:abbreviations+nomenclature}
\noindent\textbf{\underline{Abbreviations}}
\begin{table}[h!]
    \begin{tabular}{l l} 
    DC-OPF  & Direct Current Optimal Power Flow \\
    MCP     & Mixed Complementarity Problem \\
    MILP    & Mixed Integer Linear Program \\
    MPEC    & Mathematical Program with Equilibrium Constraints \\
    MPCC    & Mathematical Program with Complementarity Constraints \\
    ICC     & Information communication and control 
    \end{tabular}
    \caption{\label{table:infrastructure-models-nomenclature} Abbreviations. }
\end{table} \\

\noindent\textbf{\underline{Nomenclature}}
\begin{table}[h!]
    \begin{tabular}{l l} 
    \textbf{Definition} & \textbf{Description} \\ [0.5ex] 
    $\mathcal{N}$ & \textbf{Set} of electric power network nodes of Manhattan   \\
    $\mathcal{E}$ & \textbf{Set} of electric power network edges of Manhattan   \\
    $\mathcal{S}$ & \textbf{Set} of weather seasons   \\
    $\mathcal{H}$ & \textbf{Set} of 1-hour time steps                        \\ 
    $\mathcal{G}$ & \textbf{Set} of power generation technologies by node \\
    $\mathcal{Y}$ & \textbf{Set} of feasible space of all DC-OPF variables \\
    $\mathcal{Y}_1$ & \textbf{Set} of feasible space of all non-negative DC-OPF variables \\
    $\mathcal{Y}_2$ & \textbf{Set} of feasible space of all free DC-OPF variables \\
    $\mathcal{Y}_{1sh}^{(i)}$ & \textbf{Set} of feasible space of coordinate $i$ of set $\mathcal{Y}_1$, $i=1,2,\dots,5$ \\
    $\mathcal{Y}_{2sh}^{(i)}$ & \textbf{Set} of feasible space of coordinate $i$ of set $\mathcal{Y}_2$, $i=1,2,\dots,8$ \\
    $N=|\mathcal{N}|$ & \textbf{Parameter} of cardinality of set $\mathcal{N}$   \\
    $E=|\mathcal{E}|$ & \textbf{Parameter} of cardinality of set $\mathcal{E}$   \\       $H=|\mathcal{H}|$ & \textbf{Parameter} of cardinality of set $\mathcal{H}$   \\
    $G=|\mathcal{G}|$ & \textbf{Parameter} of cardinality of set $\mathcal{T}$   \\
    $Y=|\mathcal{Y}|$ & \textbf{Parameter} of cardinality of set $\mathcal{Y}$   \\
    $Y_1=|\mathcal{Y}_1|$ & \textbf{Parameter} of cardinality of set $\mathcal{Y}_1$   \\
    $Y_2=|\mathcal{Y}_2|$ & \textbf{Parameter} of cardinality of set $\mathcal{Y}_2$   \\
    $A\in M_{E\times N}\left(\R\right)$ & \textbf{Incidence matrix} of the electric power network \\
    $b\in \R^E$ & \textbf{Parameter} of susceptance of transmission lines \\
    $B=\text{diag}\left\{b\right\}\in M_{E\times E}\left(\R\right)$ & \textbf{Susceptance Matrix} of transmission lines \\
    $e_{i}\in\R^N$ & \textbf{Vector} of all zeros except for entry $i$ value of ``1" \\
    $z^g_{sh}\in\R^{G}$  & \textbf{Upper Level Variable} of compromised power generation capacity \\
                        & in season $s\in\mathcal{S}$, during hour $h\in\mathcal{H}$ (MWh) \\
    $z^\theta_{sh}\in\R^{N}$  & \textbf{Upper Level Variable} of compromised angle difference limit \\
                        & in season $s\in\mathcal{S}$, during hour $h\in\mathcal{H}$ (rad) \\
    $z^f_{sh}\in\R^{E}$    & \textbf{Upper Level Variable} of compromised power flow capacity in all edges \\
                        & in season $s\in\mathcal{S}$, during hour $h\in\mathcal{H}$ (MWh) \\
    $\Tilde{b}\in\R_{\geq 0}$ & \textbf{Parameter} of budget for all resources of the cyberattacker \\
    $\Tilde{c}^g_{sh}\in\R^G$  & \textbf{Parameters} of cost of compromising power generation capacity \\
                            & (proportional to budget $\Tilde{b}$) \\
    $\Tilde{c}^\theta_{sh}\in\R^E$  & \textbf{Parameters} of cost of compromising voltage angle differences \\
                            & (proportional to budget $\Tilde{b}$) \\
    $\Tilde{c}^f_{sh}\in\R^E$  & \textbf{Parameters} of cost of compromising power flow capacity \\
                            & (proportional to budget $\Tilde{b}$) \\
    \end{tabular}
\end{table} \\

\newpage
\noindent\textbf{\underline{Nomenclature (continued)}}
\begin{table}[h!]
    \begin{tabular}{l l} 
    \textbf{Definition} & \textbf{Description} \\ [0.5ex] 
    $g_{sh}\in\R^{G}$  & \textbf{Lower-Level Primal Variable} of power generation \\
    &in season $s\in\mathcal{S}$, during hour $h\in\mathcal{H}$ (MWh) \\
    $\theta_{sh}\in\R^{N}$  & \textbf{Lower-Level Primal Variable} of angle difference between nodes\\
    & in season $s\in\mathcal{S}$, during hour $h\in\mathcal{H}$ (rad) \\
    $f_{sh}\in\R^{E}$  & \textbf{Lower-Level Primal Variable} of power flow in all edges \\
                    & in season $s\in\mathcal{S}$, during hour $h\in\mathcal{H}$ (MWh) \\
    $u_{sh}\in\R^{N}_{\geq 0}$  & \textbf{Lower-Level Primal Variable} of unserved electricty demand \\
    & in all nodes and in season $s\in\mathcal{S}$, during hour $h\in\mathcal{H}$ (MWh) \\
    $y^g_{sh}\in\R^{G}$ & \textbf{Lower-Level Primal Variable} of investment \\
    & in power generation capacity in season $s\in\mathcal{S}$,  \\
    $y^f_{sh}\in\R^{E}$ & \textbf{Lower-Level Primal Variable} of investment \\
    & in power flow capacity in season $s\in\mathcal{S}$, during hour $h\in\mathcal{H}$ (MW) \\
    $\underline{g}_{sh},\bar{g}_{sh}\in\R^{G}$  & \textbf{Parameters} of lower and upper power generation limits  \\
    & in season $s\in\mathcal{S}$, during hour $h\in\mathcal{H}$ (MWh) \\
    $\underline{f}_{sh}, \bar{f}_{sh}\in\R^{E}$  & \textbf{Parameters} of lower and upper power flow limits  \\
    & in season $s\in\mathcal{S}$, during hour $h\in\mathcal{H}$ (MWh) \\
    $\underline{\theta}_{sh}, \bar{\theta}_{sh}\in\R^{E}$  & \textbf{Parameters} of lower and upper angle difference limits across all edges \\
    & in season $s\in\mathcal{S}$, during hour $h\in\mathcal{H}$ (rad)  \\
    $c^{g,l}_{sh}\in\R^{G}_{\geq 0}$    & \textbf{Parameters} of linear term of generation cost function \\
    $c^{g,q}_{sh}\in\R^{G}_{\geq 0}$    & \textbf{Parameters} of quadratic term of generation cost function \\
    $c^{f,l}_{sh}\in\R^{G}_{\geq 0}$    & \textbf{Parameters} of linear term of flow cost function \\
    $c^{f,q}_{sh}\in\R^{G}_{\geq 0}$    & \textbf{Parameters} of quadratic term of flow cost function \\
    &in season $s\in\mathcal{S}$, during hour $h\in\mathcal{H}$ (1000\$/MWh) \\
    $c^u_{sh}\in\R^{N}_{\geq 0}$    & \textbf{Parameter} of value of lost load  \\
    &in season $s\in\mathcal{S}$, during hour $t\in\mathcal{T}$ (1000\$/MWh) \\
    $d_{sh}\in\R^{N}_{\geq 0}$    & \textbf{Parameter} of total electricity demand \\
    $d^w_{sh}\in\R^{N}_{\geq 0}$    & \textbf{Parameter} of electricity demand from the water/wastewater sector \\
    & in season $s\in\mathcal{S}$, during hour $h\in\mathcal{H}$ (MWh) \\
    $d^o_{sh}\in\R^{N}_{\geq 0}$    & \textbf{Parameter} of electricity demand from all sectors excluding \\
    &  the water/wastewater sector in season $s\in\mathcal{S}$, \\
   & during hour $h\in\mathcal{H}$ (MWh)  \\ $\pi^d_{sh}\in\R^{N}$ & \textbf{Lower-Level Dual Variable} of market-clearing constraint \eqref{eq:opf-lp-mkt-clrng-cnstr} \\
& (1000\$/MWh)  \\    $\pi^f_{sh}\in\R^{E}$ & \textbf{Lower-Level Dual Variable} of power flow voltage \\
& angle constraint \eqref{eq:opf-lp-pow-flow-cnstr} (1000\$/MWh) \\    $\underline{\rho}^g_{sh},\bar{\rho}^g_{sh}\in\R^G_{\geq 0}$ & \textbf{Lower-Level Dual Variables} of power generation constraints (1000\$/MW) \\
    $\underline{\rho}^f_{sh},\bar{\rho}^f_{sh}\in\R^E_{\geq 0}$ & \textbf{Lower-Level Dual Variables} of power flow constraints (1000\$/MWh) \\
    $\underline{\rho}^\theta_{sh},\bar{\rho}^\theta_{sh}\in\R^E_{\geq 0}$ & \textbf{Lower-Level Dual Variables} of voltage angle constraints (1000\$/rad) \\ 
    $\delta_{sh}\in\R$ & \textbf{Lower-Level Dual Variable} of reference angle constraint (1000\$/rad) \\
    $C:\mathcal{Y}\rightarrow\R$  & \textbf{Function} of power system operating cost (1000\$) \\
    $\Tilde{C}:\mathcal{Y}\times\prod\limits_{sh}\R^G\times\prod\limits_{sh}\R^E\times\prod\limits_{sh}\R^E\rightarrow\R$  & \textbf{Function} of cyberattacker targeted disruption \\
    \end{tabular}
    \caption{\label{table:nomenclature} Nomenclature.}
\end{table}

\end{document}